\documentclass[aip,jcp,amsmath,amssymb,reprint,longbibliography,groupedaddress]{revtex4-1}
\usepackage[utf8]{inputenc}
\usepackage{braket}
\usepackage{amsmath}
\usepackage{bbm}
\usepackage{appendix}
\usepackage{amsfonts}
\usepackage{comment}
\usepackage{cancel}
\usepackage{bbold}
\usepackage{graphicx}
\usepackage{float}
\usepackage{MnSymbol}
\usepackage{mathtools}
\usepackage{soul}
\usepackage{tabularx}
\usepackage{array}
\usepackage{accents}
\usepackage{bm}
\usepackage{placeins}

\newcolumntype{Y}{>{\centering\arraybackslash}X}

\usepackage[colorlinks, linkcolor=blue]{hyperref}
\usepackage[nameinlink, capitalise]{cleveref}

\newcommand{\nn}{\nonumber} 
\newcommand\numeq[1]%
{\stackrel{\scriptscriptstyle(\mkern-1.5mu#1\mkern-1.5mu)}{=}}

\newcommand{\llv}{\left \lvert}
\newcommand{\rrv}{\right \rvert}
\newcommand{\tcb}{\textcolor{blue}}
\newcommand{\tcr}{\textcolor{red}}

\newcommand{\bs}{\boldsymbol}

\newcommand\commin[1]{\iffalse #1 \fi}

\newcommand{\mr}{\mathrm}

\newcommand{\mc}{\mathcal}

\newcommand{\e}{\ensuremath{\,\mathrm{e}}}

\DeclareMathOperator{\tr}{tr}

\newcommand{\ba}{{\mathrm{b}}}
\newcommand{\sy}{{\mathrm{s}}}
\newcommand{\syba}{{\mathrm{sb}}}

\newcommand\norm[1]{\left\lVert#1\right\rVert}

\usepackage{chngcntr}

\numberwithin{equation}{section}

\begin{document}
\title{Dynamical signatures of non-Markovianity in a dissipative--driven qubit}
\author{Graziano Amati}
\altaffiliation{Present Address: Frick Chemistry Laboratory, Princeton University, Princeton, NJ 08544, United States.}
\affiliation{Albert Ludwigs Universit\"at Freiburg, Hermann Herder Str. 3, 79104 Freiburg, Germany\looseness=-1}

\date{\today}

\begin{abstract}
We investigate signatures of non-Markovianity in the dynamics of a periodically-driven qubit coupled to a dissipative bosonic environment.
We propagate the dynamics of the reduced density matrix of the qubit by integrating the numerically exact hierarchical equations of motion.
Non-Markovian features are quantified by comparing on an equal footing the predictions from diverse and complementary approaches to quantum dissipation.
In particular, we analyze the distinguishability of quantum states, 
the decay of the volume accessible to the qubit on the Hilbert space, the negativity of the canonical rates in the generalized Lindblad equation and the relaxation of the memory kernels in the Nakajima--Zwanzig generalized quantum master equation.
We study the effects of controlled driving on the coherent dynamics of the system.
We show that a suitable external field can offset the ergodic relaxation of time correlation functions, increase distinguishability over time and strengthen non-Markovian effects witnessed by the canonical dissipation channels.
We furthermore observe the phenomenon of eternal non-Markovianity for sufficiently small system--bath coupling and we discuss how this can be enhanced by modulating the frequency of the external drive.
The present work provides a broad theoretical analysis of quantum dissipation in the framework of open quantum dynamics and quantum information. 
\end{abstract}

 \maketitle

\section{Introduction}


To date, uncontrolled dissipation is a crucial issue limiting both the reliability of quantum-information protocols and the scalability of quantum-computing devices.
\cite{babu2021,ray2023,verney2019,harrington2022} 
Improving coherence lifetime is a task of the utmost importance in order to enable quantum computers to tackle relevant and challenging numerical analyses, e.g. in the framework of drug discovery, \cite{outeiral2021} artificial intelligence \cite{ahmed2021} and the simulation of many-body quantum systems. \cite{smith2019,franke2019}

From a theoretical standpoint, it is in the first place controversial how to systematically define and measure dissipation in quantum mechanics.
It is in particular debated how this ubiquitous physical process relates to the occurrence of memory effects, otherwise called signatures of \emph{non-Markovianity} (in contrast to Markovian memory-less processes). \cite{breuer2012,rivas2014}
\\
Memory effects in classical dynamics can be witnessed by measuring the probability of a physical system of being in a given state in the present, conditioned on its past history. \cite{breuer2016} 
This probabilistic approach is applicable provided a series of measurements is performed over time. 
While measurement processes do not alter classical states, they do however affect coherence in quantum systems.
In particular, projective measurements lead to the irreversible collapse of entangled quantum states. \cite{schlosshauer2005}
Hence, in order to minimize external perturbations induced by measurement processes, alternative methods have been proposed to quantify dissipation and non-Markovianity in quantum mechanics.

The \emph{trace-distance approach} to quantum non-Markovianity measures dissipation from the loss of distinguishability over time between two maximally distinguishable states of a quantum subsystem, due to the interaction with an external environment.\cite{breuer2009,breuer2016,amato2018} 
In particular, one can prepare the subsystem  in two states maximizing an operator metric known as the trace distance. 
Unitary 
dynamics conserve this distance, while it decreases monotonically over time in the opposite limit of strongly dissipative dynamics.
It can however occur that, even in the limit of an infinitely large dissipative bath, the trace distance increases over time. 
Increasing distinguishability has been related to the occurrence of information backflow from the environment. \cite{breuer2016,chakraborty2018} 
From this perspective, it has been proposed to leverage non-Markovianity itself as a resource for quantum information and control.  \cite{reich2015,pineda2016,mirkin2019}

The \emph{volume of states} accessible by the qubit provides another measure of dissipation induced by the environment. \cite{lorenzo2013}
It has been discussed in the literature that a drawback of this method is its potential inability to accurately capture weak dissipation occurring across a subset of channels in the Hilbert space of the subsystem.
This issue is however resolved by analyzing the \emph{canonical rates}, time-dependent coefficients appearing in an exact time-local master equation 
in Lindblad form. \cite{hall2014,andersson2007,maldonado-mundo2012,hayden2022}
The latter approach is capable of quantifying memory effects occurring along each quantum channel.
Specifically, a generalized Lindblad equation, uniquely expressed in canonical form, involves a set of time-dependent canonical dissipation rates, which are non-negative functions of time under Markovian dynamics. 
Therefore, the negativity of each rate serves as an indicator of non-Markovian behavior within the corresponding channel. \emph{Eternal non-Markovianity} arises whenever certain rates assume negative values at all times. \cite{shrikant2022,gulacsi2023}

The \emph{Nakajima--Zwanzig formalism} is another rigorous, first-principled approach to quantum dissipation. \cite{nakajima1958,zwanzig1960,mori1965} 
The method relies on expressing open quantum dynamics in terms of an exact integro--differential equation of motion.
An integral term featuring a memory kernel encapsulates complete information on the interaction between the subsystem and the environment.
The kernel vanishes in the limit of unitary dynamics, whereas it collapses to a Dirac delta function in the opposite Markovian limit.
As the timescale to relaxation of the kernel is highly sensitive to environmental dissipation and decoherence, it represents a natural measure of non-Markovianity. \cite{cohen2011,ng2022}


All aforementioned methods provide an interpretation of dissipation from different perspectives, not necessarily in agreement with each other\cite{chruscinski2011,chruscinski2017,bhattacharya2017,chanda2016} 
and, as discussed in Ref.~\onlinecite{pollock2018,pollock2018nm}, they do not recover the classical definition of Markovianity expressed by the Kolmogorov conditions.
It is still subject for debate whether a unique measure to dissipation can be defined in the first place, or whether several different approaches should be regarded as complementary to each other.

The effects of dissipation on the long-time limits of open dynamical systems can be inspected from the perspective of ergodic theory.
From a classical standpoint, environmental dissipation
leads to the ergodic relaxation of the initial distribution to its invariant thermal limit, fulfilling detailed balance.\cite{hawkins2021}
For quantum mechanical systems it has been conjectured, within the \emph{eigenstate thermalization hypothesis}, that thermalization can ultimately occur if the asymptotic limit of quantum operators 
is consistent with the microcanonical ensemble. \cite{deutsch2018,dalessio2016} 
Predicting the occurrence of chaos and thermalization in quantum dynamics remains however a hard and unresolved theoretical problem.
Analyzing the long-time behavior of open quantum dynamics becomes even more challenging in presence of an external time-dependent field.
This scenario is particularly relevant from the perspective of quantum error correction (QEC), given that coupling dissipative dynamics to an external drive is a well-know and effective strategy to protect the dynamics of the system from environmental decoherence.
 \cite{jirari2006,jirari2007,jirari2020,rebentrost2009}
In particular, loss of coherence in a system of qubits occurs naturally due to the uncontrolled coupling to an external environment.
To counterbalance the effects of environmental noise, a variety of QEC techniques has been developed in the last decades. \cite{lidar2013,vozhakov2022,terhal2015,babu2021}
Among those, a successful family of approaches is based on \emph{dynamical-decoupling techniques}, protecting the dynamics from environmental dissipation by means of controlled external pulses. 
\cite{bastrakova2019,rebentrost2009,viola1999,viola2000}

Inspired by dynamical-decoupling strategies, in the present work we analyze how an external time-dependent drive affects environmental dissipation in a spin--boson model externally coupled to a periodic monochromatic field. \cite{engelhardt2019,magazzu2018exp}
We propagate the open quantum dynamics of the qubit by integrating the  hierarchical equations of motion (HEOM). \cite{tanimura2014,tanimura2020}
This allows us to obtain a numerically exact solution for the reduced density matrix (RDM) of the qubit.

In \cref{subsec:dyn_ergo} we study how the ergodic relaxation of a complete set of time correlation functions is affected by the presence of the external drive.
In \cref{subsubsec:TD} we calculate a witness of non-Markovianity based on the distinguishability of quantum states, within the trace-distance formalism. \cite{breuer2009}
From our numerical analysis it emerges that the onset of environmental dissipation can be minimized provided the quantum system is periodically driven at given resonant frequencies. \cite{shuang2000,mori2023} 
After introducing in \cref{subsubsec:volume} the analysis of the volume of accessible states and related drawbacks, we show in  \cref{subsubsec:canonical} that controlled driving can be tuned to measure the onset of stable and long-lived non-Markovian effects. 
This analysis is developed within the framework of an exact generalized Lindblad master equation.
Despite its efficacy, the study of the canonical rates is exclusively applicable provided the map generating the open quantum dynamics is invertible.
This issue does not affect the measurement of non-Markovian effects from the timescale to relaxation of the kernels of the generalized quantum master equation (GQME) in the Nakajima--Zwanzig formalism.\cite{nakajima1958,zwanzig1960} We study this approach in \cref{subsubsec:GQME}.

This work provides an extensive analysis of quantum non-Markovianity, and paves the way towards the extended study of dissipation in large-scale quantum-information systems.

\section{Dissipative--driven dynamics}\label{sec:diss_drive}

In the present work we study the dynamics of a fermionic two-level  subsystem ($\mr s$) coupled to a bosonic environmental bath ($\mr b$) and to an external time-dependent drive. 
The total Hilbert space $\mc H$ can be factorized in terms of the tensor product of the subsystem and bath Hilbert spaces, i.e. $\mc H=\mc H_{\mr s}\otimes \mc H_{\mr b}$ .
The Hamiltonian is decomposed onto three contributions, according to
\begin{equation}\label{eq:Hgen}
\hat H(t) = \hat H_{\mr s}(t) \otimes \hat {\mc I}_{\mr b}+ \hat {\mc I}_{\mr s}\otimes \hat H_{\mr b} + \hat H_{\mr {sb}}.
\end{equation}
Here, we denote by $\hat H_{\mr s}(t) $ the Hamiltonian of the qubit subsystem, which is coupled to the bosonic bath $\hat H_{\mr b}$ via the cross term $\hat H_{\mr {sb}}$.
$\hat {\mc I}_{\mr s}$ and $\hat {\mc I}_{\mr b}$ denote the identity operators of the subsystem and the bath, respectively.
The time evolution of the total density matrix $\hat\rho_0$ is expressed via the time-ordered exponential
\begin{subequations}\label{eq:tot_prop}
\begin{align}
\hat\rho(t,\tau) &= \exp_+\left\{-i\int_\tau^t\mr d t'\; \mc L_{t'}\right\} \hat\rho_0\\
&= \hat U(t,\tau)\hat\rho_0\hat U^\dagger(t,\tau),
\end{align}
\end{subequations}
where 
$
\mc L_t \;\cdot = [\hat H(t),\cdot]
$
 denotes the Liouville superoperator, the subscript $+$ indicates positive time ordering in the exponential superoperator  
and
\begin{equation}\label{eq:U}
\hat U(t,\tau) =  \exp_+\left\{-i\int_\tau^t\mr d t'\; \hat H(t')\right\}
\end{equation}
is the time evolution operator.
Here and in the following we fix $\hbar=1$.

The open dynamics of the subsystem are fully described by the RDM $\hat\rho_{\mr s}(t,\tau) = \tr_{\mr b}\{\hat\rho(t,\tau)\}$, where $ \tr_{\mr b}\{\cdot\}$ denotes the partial trace over the Hilbert space of the bath.
The total trace $\tr[\cdot ] $ is obtained by composing ${\tr[\cdot ] =\tr_{\mr s}[\tr_{\mr b}\{\cdot\} ] = \tr_{\mr b}\{\tr_{\mr s}[\cdot]\}}$, where $\tr_{\mr s}[\cdot ]$ is the partial trace with respect to the subsystem.
We define $\hat\rho(t)=\hat\rho(t,0)$ and similarly for
the time evolution of other operators and superoperators.

In the present work we consider an initially factorized total density matrix of Feynman--Vernon type \cite{feynman1963}
\begin{equation}\label{eq:rho0}
\hat \rho(0) = \hat \rho_0 = \frac 1 2\left(v_i \hat \sigma_i+\hat{\mc I}_{\mr s}\right)\otimes \hat\rho_{\mr b} = \frac 1 2v_\mu\hat\sigma_\mu\otimes\hat\rho_{\mr b},
\end{equation}
where
\begin{equation}
\hat\rho_{\mr b}=\frac 1 {Z_{\mr b}}\e^{-\beta \hat H_{\mr b}},\hspace{10mm} Z_{\mr b} = \tr_{\mr b}\{\e^{-\beta \hat H_{\mr b}}\},
\end{equation}
denotes the thermal density matrix for the uncoupled bath.
Throughout the paper we make use of Einstein's summation convention, reserving Latin letters $i,j,k,\cdots \in\{x,y,z\}$ for sums over the indices of the three Pauli matrices $\hat \sigma_i$'s.
We instead utilize Greek letters $\mu,\nu,\lambda,\cdots\in\{0,x,y,z\}$ to include in the sum the identity operator of the qubit subsystem $\hat \sigma_0 = \hat{\mc I}_{\mr s}$.
This quadruplet forms a basis for the Hilbert space of the qubit.
The coefficients $v_\mu$'s of the Bloch vector can be chosen to be real without loss of generality.
Finally, the normalization condition $\tr\{\hat\rho(t)\}=1$ holds by fixing $v_0=1$.
\\

We discuss in the following different formulations of the open dynamics of the qubit, and the mutual relations between those approaches.

The dynamics of the RDM can be written under broad conditions in terms of the Nakajima--Zwanzig generalized quantum master equation (GQME) \cite{nakajima1958,zwanzig1960} 
\begin{subequations}\label{eq:GQME_RDM}
\begin{align}
\frac{\mr d}{\mr d t}\hat\rho_{\mr s}(t) &= - i[\hat H_{\mr s}(t), \hat\rho_{\mr s }(t)]- \int_0^t \mr d \tau\; k_{t,\tau}\hat \rho_{\mr s}(\tau)\label{eq:GQME_RDM1}\\
&= \Lambda_t \hat\rho_{\mr s}(t).\label{eq:GQME_RDM2}
\end{align}
\end{subequations}
Here, $k_{t,\tau}$ denotes the kernel superoperator, 
\begin{equation}\label{eq:Lambda}
\Lambda_t \bullet = - i [\hat H_{\mr s}(t), \cdot] - \int_0^t\mr d \tau\; k_{t,\tau} \circ \Phi_{\tau}\Phi_{t}^{-1}\bullet
\end{equation}
is the generator of the open dynamics of $\hat\rho_{\mr s}(t) $ and $\Phi_t $ the corresponding propagator
\begin{equation}\label{eq:Phi}
\Phi_t \;\bullet = \tr_{\mr b}\left\{\exp_+\left\{- i\int_0^t \mr d t'\; \mc L_{t'}\right\}(\bullet \otimes \hat \rho_{\mr b})\right\}.
\end{equation}
\Cref{eq:GQME_RDM} is solved by
\begin{equation}
\hat\rho_{\mr s}(t) = \exp_+\left\{\int_0^t\mr d t'\; \Lambda_{t'}\right\}\hat\rho_{\mr s}(0) =\Phi_t \hat\rho_{\mr s}(0) .
\end{equation}
A derivation of \cref{eq:GQME_RDM} for time-correlation functions is provided in \cref{app:GQME}. 
In the Born--Markov approximation, the memory kernel is replaced by a Dirac delta function [i.e. $k_{t,\tau} \propto \delta(t-\tau)$].
We will discuss in \cref{subsubsec:GQME} how deviations from this limit can be utilized as a measure of non-Markovian effects.
\\
From \cref{eq:GQME_RDM,eq:Lambda} it is evident that a time-local representation of the dynamics is possible provided the propagator $\Phi_t$ is invertible, while the GQME \cref{eq:GQME_RDM2} is not affected by this restriction.
If $\Phi_{t}^{-1}$ exists at a given time $t$, it \tcr{is} possible to construct a time-local master equation in 
canonical form \cite{hall2014,andersson2007,maldonado-mundo2012,hayden2022}
\begin{align}
\Lambda_t{\hat\rho}_{\mr s}(t) &= -i [\hat H_{\mr c}(t),\hat \rho_{\mr s}(t)] \nn \\
&\quad + \gamma_i(t) \left[\hat L_i(t) \hat\rho_{\mr s}(t)\hat L^\dagger_i(t)  -\tfrac 12 \{\hat L_i^\dagger(t)\hat L_i(t), \hat\rho_{\mr s}(t)\} \right].\label{eq:canonical}
\end{align}
\Cref{eq:canonical} is expressed in terms of a uniquely defined set of canonical rates $\gamma_i(t)$'s and an orthogonal basis of traceless 
operators $\hat L_i(t)$'s, fulfilling
\begin{equation}\label{eq:traceL}
\tr_{\mr s}[\hat L_i(t)] =0, \hspace{10mm} \tr_{\mr s}[\hat L_i(t)\hat L_j(t)] = 2\delta_{ij}.
\end{equation}
$\hat H_{\mr c}(t)$ is the canonical ($\mr c$) Hamiltonian generating the Hermitian part of the dynamics.
A derivation of \cref{eq:canonical} is given in \cref{app:canonical}.
According to the \emph{Gorini--Kossakowski--Sudarshan theorem}, the non-negativity of the rates $\gamma_i(t)$'s is a  necessary and sufficient condition for the complete positivity of the dynamics. \cite{gorini1976,breuer2016}
In \cref{subsubsec:canonical} we will exploit this feature to measure non-Markovianity in open quantum dynamics with the only --- albeit strong --- restriction to invertible dynamical maps.
\\

The spin time-correlation functions (STCF)
\begin{align}\label{eq:Cmn}
\mc C_{\mu\nu}(t) &= \frac 1 2 \tr\left[(\hat\sigma_\mu\otimes\hat\rho_{\mr b})\exp_{-}\left\{i\int_0^t\mr d t'\; \mc L_{t'}\right\}\hat\sigma_\nu\right]
\end{align}
allow to naturally decompose the contributions of populations and coherences in the dynamics, and to conveniently measure the invertibility of the dynamical map.
The time evolution of \cref{eq:Cmn} provides a description of the dynamics equivalent to the RDM.
In fact, we can rewrite
\begin{align} 
\hat\rho_{\mr s}(t) 
&=\frac 12 \tr[\hat \rho(t)\hat \sigma_\nu]\hat\sigma_\nu = \frac 1 4 \tr[(v_\mu\hat\sigma_\mu\otimes \hat \rho_{\mr b} )\hat \sigma_\nu(t)]\hat\sigma_\nu\nn\\
&= \frac 1 2 v_\mu \mc C_{\mu\nu}(t)\hat\sigma_\nu. \label{eq:rhos_C}
\end{align}
An important property of the matrix of the STCF [with components \cref{eq:Cmn}] is that its transposed corresponds to the propagator of the Bloch vector.
This can be shown by expanding
\begin{equation}\label{eq:rhos_v}
\hat\rho_{\mr s}(t)  
=\frac 12 v_\mu(t) \hat\sigma_\mu,
\end{equation}
and by making use of \cref{eq:Phi}, to obtain 
\begin{align}
v_\mu(t) &= \frac 12  \tr_{\mr s}\left[\hat \sigma_\mu\Phi_t [\hat \sigma_\nu]\right] v_\nu \nn\\
&=\frac 12 \tr_{\mr s} \left[(\hat \rho_{\mr b}\otimes\hat \sigma_\nu)  \tr_{\mr b} \left\{\exp_-\left\{\frac i \hbar \int_0^t \mr d t'\; \mc L_{t'}\right\}\hat\sigma_\nu\right\}\right] v_\nu \nn\\
&= \mc C^T_{\mu\nu} (t) v_\nu. \label{eq:bloch_t}
\end{align}
The norm of the Bloch vector is conserved by the dynamics of an isolated subsystem, while it decreases monotonically for a strongly dissipative bath. \cite{breuer2009,breuer2016}  
We will utilize these geometric features to witness non-Markovian effects in the later \cref{subsubsec:TD}.

Let us remark that the present formalism can be straightforwardly generalized to a subsystem consisting of an arbitrary number $N$ of quantum levels.
In particular, one can replace the Pauli matrices with a set of $N^2-1$ operators $\{\hat\Sigma_i\}_{i=1}^{N^2-1}$ generating the $\mathfrak{su}(N)$ algebra of the subsystem and obeying the trace relations \cite{runeson2020,lorenzo2013volume,kimura2003}
\begin{equation}\label{eq:genSigma}
\tr_{\mr s}[\hat \Sigma_i \hat \Sigma_j] = N \delta_{ij}, \hspace{10mm}\tr_{\mr s}[\hat \Sigma_i]=0.
\end{equation}
The $\Sigma_i$'s, together with the $N\times N$ identity operator, constitute a basis of the Hilbert space of the subsystem.
\\

In \cref{sec:model} we present in detail the model system studied in this work, while in \cref{sec:num} we develop an extensive numerical analysis of quantum non-Markovian effects from multiple perspectives.

\section{Model system} \label{sec:model}
A qubit bilinearly coupled to a bosonic field and to a time-dependent pulse is modeled by a periodically driven spin--boson model. \cite{grifoni1999,magazzu2018exp,magazzu2018theo, cangemi2019}
We consider here an external drive exclusively coupled to the Hamiltonian of the subsystem, leading to periodic oscillations in the population difference.
Specifically, the different terms of \cref{eq:Hgen} are  
\begin{subequations}\label{eq:Hspec}
\begin{align}
\hat H_\sy(t) &= \Delta\hat\sigma_x+[\varepsilon_0 + \varepsilon_d\cos(\Omega t)]\hat\sigma_z, \label{eq:Hs}\\
\hat H_\ba &= \sum_{\alpha=1}^F\omega_\alpha\hat a^\dagger_\alpha\hat a_\alpha, \label{eq:Hb}\\
\hat H_\syba &= \hat\sigma_z\otimes  \sum_{\alpha=1}^F c_\alpha (\hat a^\dagger_\alpha+\hat a_\alpha).\label{eq:Hsb}
\end{align}
\end{subequations} 
The diagonal part of $\hat H_{\mr s}(t)$ includes a static bias $\varepsilon_0$ and a time-dependent term involving periodic oscillations at frequency $\Omega$ and amplitude $\varepsilon_{\mr d}$.
The two quantum levels interact through the coupling $\Delta \hat \sigma_x$, where $\Delta$ denotes the state--state coupling constant.
$\hat a_\alpha^\dagger$'s ($\hat a_\alpha$'s) are the creation (destruction) operators of the $F$ bath modes, oscillating at frequencies $\omega_\alpha$'s and coupled to the population difference via the interaction defined by the coefficients $c_\alpha$'s.
In this work we consider an Ohmic spectral density with a Drude cutoff \cite{weiss2012}
\begin{equation}\label{eq:J}
	J(\omega) = \frac{\eta}{\pi}\frac{\omega_{\mr c}\omega}{\omega_{\mr c}^2+\omega^2}.
\end{equation}
$\eta$ and $\omega_{\mr c}$ in \cref{eq:J} denote respectively the system--bath coupling constant and the cutoff frequency of the bath modes.
\Cref{eq:J} corresponds to the continuous limit of its discrete counterpart
\begin{equation}\label{eq:JF} 
	J_F(\omega) = \sum_{\alpha=1}^F \frac{c_\alpha^2}{2m_\alpha\omega_\alpha}\delta(\omega-\omega_\alpha).
\end{equation}
The degrees of freedom of the bath can be rewritten in terms of canonical configurations $\hat q_\alpha$'s and momenta $\hat p_\alpha$'s via the change of coordinates
\begin{equation}\label{eq:destrB}
\hat a_\alpha = \frac 1 {\sqrt{ 2\omega_\alpha}}\left(\sqrt{m_\alpha}\omega_\alpha\hat q_\alpha+i\frac{\hat p_\alpha}{\sqrt{m_\alpha}}\right),
\end{equation}
such that
\begin{subequations}
\begin{align}
\hat H_{\mr b}
&= \frac 1{2}\sum_{\alpha=1}^F\left(m_\alpha \omega_\alpha^2\hat q_\alpha^2+\hat p_\alpha^2\right),\\
\hat H_{\mr{sb}} &= \hat\sigma_z \otimes\sum_{\alpha=1}^F\tilde c_\alpha \hat q_\alpha,\label{eq:HSB}
\end{align}
\end{subequations}
where we made use of the canonical commutation relations $[\hat q_\alpha,\hat  p_{\alpha'}]= i\delta_{\alpha\alpha'}$.

\section{Numerical analysis }\label{sec:num}

\subsection{Dynamics and stationary state}\label{subsec:dyn_ergo}
In this section we study the time evolution of the STCF [defined in \cref{eq:Cmn}], and their asymptotic behavior at long times.

\begin{figure*}
\centering\includegraphics[width=5in]{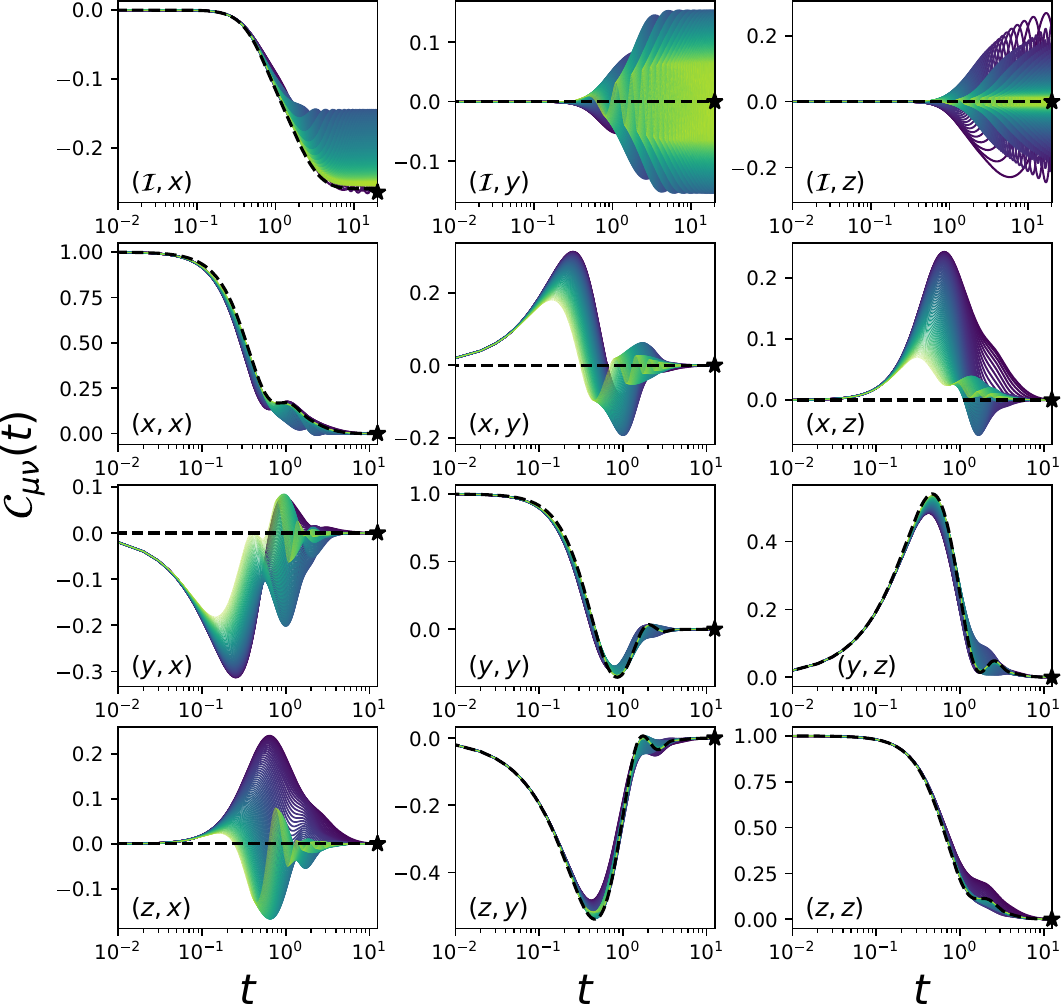}\caption{Dynamics of the spin time-correlation functions $\{\mc C_{\mu\nu(t)}\}_{\mu\nu}$ [\cref{eq:Cmn}], shown for increasing values of the frequency of the external drive from $\Omega=0$ (blue/dark) to $\Omega=20$ (yellow/light).  
The other parameters are set to $\beta=0.3$,  $\varepsilon_0=0$, $\varepsilon_{\mr d}=1$, $\eta=1$ and $\omega_{\mr{c}}=1$. The dashed lines correspond to the dynamics of the time-independent case with $\varepsilon_{\mr d}=0$ (see discussion in \cref{subsec:dyn_ergo}).
The black stars shown on all panels are the long-time benchmarks for the non-driven system in the assumption of ergodic dynamics.}\label{fig:Cmn}
\end{figure*}

Given that the identity operator $\hat \sigma_0$ is invariant under the action of the total propagator [\cref{eq:tot_prop}], $\mc C_{\mu 0}(t) = \tr[\hat\rho_0\hat\sigma_\mu \hat\sigma_0(t)]= \delta_{\mu 0}$. 
The time evolution of the other nontrivial correlations is shown in \cref{fig:Cmn} for increasing values of the frequency of the external drive from $\Omega=0$ (blue/dark) to $\Omega=20$ (yellow/ light);
the other system parameters are set to $\beta=0.3$, 
$\varepsilon_0=0$, $\varepsilon_{\mr d}=1$, $\eta=1$ and $\omega_{\mr{c}}=1$.
Also, we fix in the following $m_\alpha=1$  for all ${\alpha=1,\cdots,F}$. 
All the physical constants are given here in units of the state-state coupling constant $\Delta$, effectively set equal to 1. \cite{runeson2019,montoya-castillo2016}


We calculate the dynamics STCF's by integrating the numerically exact hierarchical equations of motion (HEOM).
These are a set of coupled differential equations for the RDM of the subsystem and additional auxiliary densities. \cite{ishizaki2005,tanimura2020}
Although this hierarchy is in principle infinite, it can be safely truncated to a finite tier for all the systems considered in this work, to obtain a numerically exact solution of the open dynamics of the qubit.
The integration of the HEOM requires to fix two numerical parameters, i.e., the number of Matsubara modes $K$ utilized in the high-temperature expansion of the bath correlation function, and the highest tier $L$ considered in the hierarchy.
All the results presented in the paper exhibit a satisfactory convergence with $K\le 2$ and $L\le 20$. 

We could ask ourselves whether any statement can be made on the long-time dynamics of the STCF shown in \cref{fig:Cmn}. 
From numerical observaions in a variety of systems and parameter regimes, \cite{ellipsoid,GQME,thermalization,CHIMIA} we consistently observed that, for a time-independent energy bias [$\varepsilon(t)=\text{const}$], the quantum generalization of the classical strong-mixing condition \cite{hawkins2021} 
\begin{equation}\label{eq:C_tindep}
\mc C_{\mu\nu}(t\to+\infty) = \langle \hat \sigma_\mu \rangle_0 \langle \hat\sigma_\nu \rangle_\beta
\end{equation}
holds at the long-times.
Here,
\begin{align}\label{eq:sigma_mu}
	\langle \hat\sigma_\mu \rangle_0 &=\frac 12 \tr[\hat \sigma_\mu \otimes \hat \rho_{\mr b}] =  \frac 12\tr_{\mr s}[\hat \sigma_\mu] = \delta_{\mu 0}
\end{align}
denotes the average of the static spin operator over the initial distribution $\hat \rho_0 = \frac 1 2 \hat{\mc I}_{\mr s}\otimes \hat\rho_{\mr b}$, while $\langle \hat \sigma_\nu\rangle_\beta$ is the average of the time-evolved operator over the thermal density matrix $\hat \rho_\beta = \e^{-\beta \hat H}/\tr[\e^{-\beta\hat H}]$.
\Cref{eq:C_tindep} is supported by the long-time limit of the black curves shown in \cref{fig:Cmn} (where $\varepsilon_{\mr d}=0$).
In fact, these agree satisfactory agree with the predictions of \cref{eq:C_tindep}, shown as black stars at the final time in all panels.
The thermal averages have been calculated by making use of the reaction-coordinate representation of the spin--boson Hamiltonian. \cite{thoss2001,wang2017} The approach is exact for an arbitrary quantum bath linearly coupled to the subsystem. Nevertheless, for our parameter regime it is possible to approximate the quantum environment with its classical counterpart. 
As discussed in Appendix B of Ref.~\onlinecite{ellipsoid}, this allows us to replace the trace over the reaction coordinate with a one-dimensional phase-space integral.
The good agreement between long-time dynamics and statistics supports the validity of both the mixing assumption for a time-independent Hamiltonian and of the classical limit for the environmental bath (as expected here, given that  $\beta\omega_{\mr c}<1$\cite{shi2009,zhu2013}). 
A generalization of \cref{eq:C_tindep} in presence of an external drive is 
\begin{equation}\label{eq:C_tdep}
\mc C_{\mu\nu}(t) \sim \langle \hat \sigma_\mu \rangle_0 \langle \hat\sigma_\nu \rangle_{t,\mr{st}},\hspace{10mm} t\gg 1.
\end{equation}
Here, 
\begin{equation}\label{eq:sig_mu0}
	\langle \hat\sigma_\nu \rangle_{\mr{st}, t} = \tr[\hat \rho_{\mr{st}}(t)\hat\sigma_\nu]
\end{equation}
is the average over the stationary out-of-equilibrium density matrix $\hat \rho_{\mr{st}}(t)$ which, according to the Floquet theorem, obeys the same periodicity of the driven Hamiltonian \cref{eq:Hgen}. \cite{sato2020,ikeda2020}
To the best of our knowledge, an exact expression of $\hat\rho_{\mr st}(t)$ has not been provided in the literature thus far. 
Approximations of this stationary density have at least been derived in certain limits, e.g. strong but non-vanishing drive or small system--bath coupling. \cite{shirai2016,engelhardt2019} 
For our model system we can show a necessary condition for \cref{eq:C_tdep} to be valid in presence of a time-dependent drive, i.e. that $\mc C_{i\nu}(t\to \infty)=0$ for $i\neq 0$, as expected from \cref{eq:sigma_mu}.

By inspecting \cref{fig:Cmn}, we note that, as we increase the frequency $\Omega$ (from blue to red), the driven dynamics converge to the time-independent limit, with $\varepsilon_{\mr d}=0$ (black dashed lines).
This ``washout'' effect can be understood by noticing that the dynamics generated by \cref{eq:Hgen,eq:Hspec} depend on the drive exclusively via the factor $\varepsilon_{\mr d}\sin(\Omega t)/\Omega$ (see \cref{app:rotframe} for a proof).   
This implies that $\Omega\to+\infty$ and $\varepsilon_{\mr d}\to 0$ are equivalent limits from a dynamical standpoint.



A convenient strategy for reducing dissipation in open quantum dynamics is to maximize non-Markovian effects. In fact, strongly dissipative processes tend to display Markovian memory-less features.
Non-Markovian effects are instead associated to a weak-to-intermediate coupling between a quantum subsystem and an external environment (see the later discussion in \cref{subsubsec:GQME}).
In the following \cref{subsec:nonmark} we discuss the efficacy of several approaches to quantum non-Markovianity, and the involved interplay between environmental dissipation and controlled drive.

\subsection{Non-Markovianity}\label{subsec:nonmark}
\subsubsection{Trace distance}\label{subsubsec:TD}
The trace-distance approach to quantum non-Markovianity relies on the interpretation of dissipation as loss of distinguishability between quantum states over time. \cite{vacchini2012,breuer2016}
In this section we outline the fundamental aspects of this approach, and we apply it to the present study of dissipative--driven dynamics. 
Extended discussions and reviews on the method can be found, e.g., in Refs.~\onlinecite{breuer2010,breuer2016}.
\\

Let us consider two copies of the qubit prepared in two well-defined initial states, with RDM's $\hat \rho_{\mr s}^{(1)}(0)$ and $\hat \rho_{\mr s}^{(2)}(0)$ and corresponding Bloch vectors $\bm v^{(1)}(0)$ and $\bm v^{(2)}(0)$.
A measure of distinguishability between these states is given by the \emph{trace distance} 
\begin{equation}\label{eq:Dab}
\mc D^{(1,2)}(t) = \frac 12\norm{\hat\rho_{\mr s}^{(1)}(t)-\hat\rho_{\mr s}^{(2)}(t)},
\end{equation}
where $\norm{\cdot}$ is the trace norm on $\mc H_{\mr s}$. \cite{watrous2018}
Not that the trace distance of a Hermitian operator $\hat A\in\mc H_{\mr s}$ can be expressed in terms of the absolute sum of its eigenvalues $d_i$'s, that is $\norm{\hat A} =\sum_{i=1,2,3}|d_i|$. \cite{breuer2016} 
This, in particular, implies that for a two-level subsystem the trace distance coincides with half of the Euclidean distance between the respective Bloch vectors:
\begin{equation} 
\mc D^{(1,2)}(t) = \frac 12 |\bm v^{(1)}(t)-\bm v^{(2)}(t)|.
\end{equation}
Equivalently, by making use of \cref{eq:bloch_t},
\begin{align}\label{eq:D_v}
\mc D^{(1,2)}(t) 
&=\frac 12\left\{\sum_{j=x,y,z} \left[\sum_{i=x,y,z}\left(v^{(1)}_i-v^{(2)}_i\right)\mc C_{ij}(t)\right]^2\right\}^{1/2}.
\end{align}
Let us note that the correlations $\mc C_{0\nu}(t)$, $\nu\in\{0,x,y,z\}$ do not contribute to the trace distance, and that all the terms involved in \cref{eq:D_v} relax to zero at long times (as discussed in \cref{subsec:dyn_ergo}).
This indicates that information on the initial state is completely dissipated to the environment at long times.
A limitation of the trace distance is that the approach cannot capture non-Markovian effects due to correlation functions $\mc C_{0\nu}(t)$ generating the non-unital part of the dynamics and leading to nonzero stationary limits. \cite{utagi2023}

\begin{figure*}[!htb]
\begin{minipage}[t]{0.49\textwidth}
\centering\includegraphics[width=1.1\linewidth]{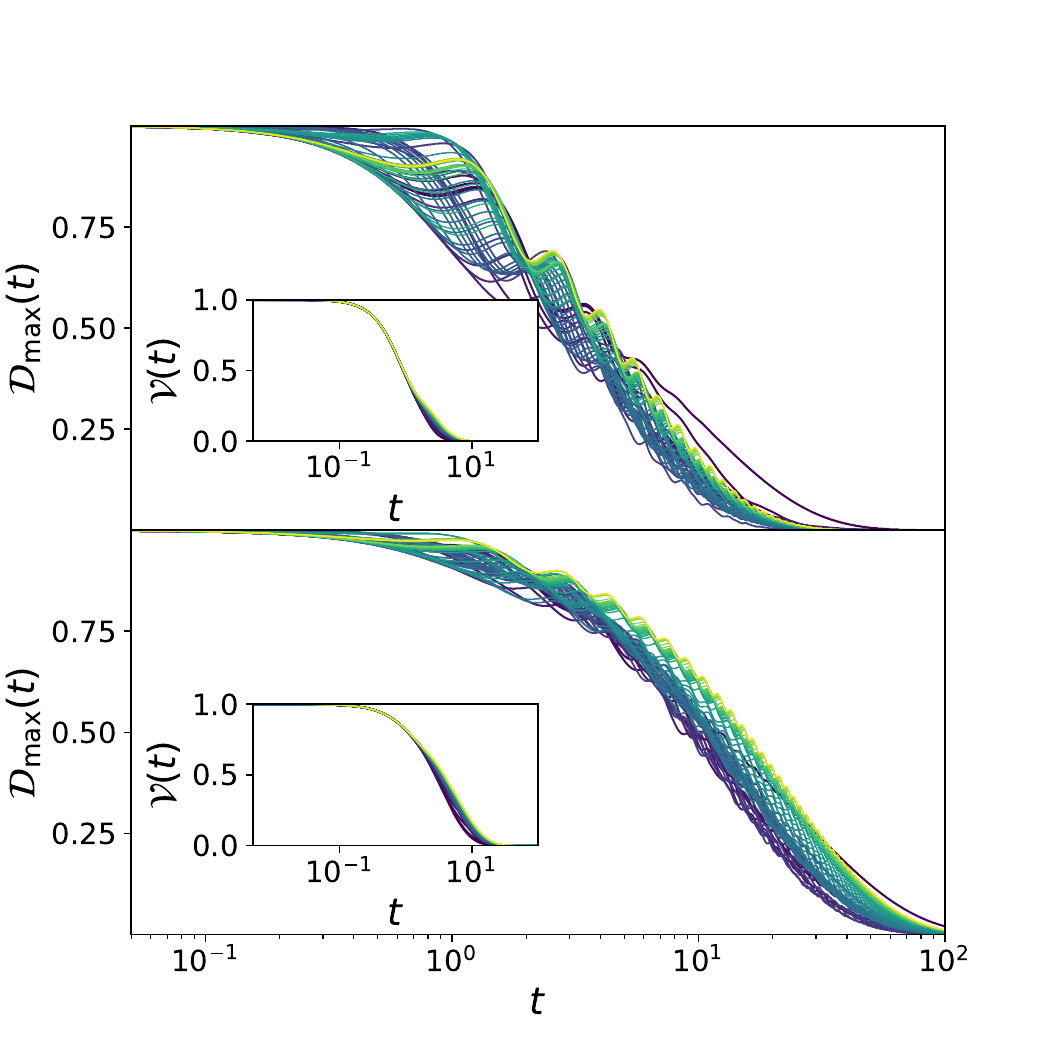}\caption{Main panels: time evolution of the maximal trace distance $\mc D_{\mr{max}}(t)$ for the dissipative--driven qubit.
Results are shown for two values of the inverse temperature, $\beta=0.3$ and $\beta=1.6$ and for increasing values of the frequency of the drive from $\Omega=0$ (blue/dark) to $\Omega=20$ (yellow/light). 
The other parameters are set to $\varepsilon_0=0$, $\varepsilon_{\mr d}=1$, $\eta=0.1$ and $\omega_{\mr c}=1$. Insets: time evolution of the volume of accessible states, with the same color code and for the same systems shown in the main panels (see the discussion in \cref{subsubsec:volume}.) }
\label{fig:Dmax0.1}
   \end{minipage}\hfill
   \begin{minipage}[t]{0.49\textwidth}
    \centering\includegraphics[width=1.1\linewidth]{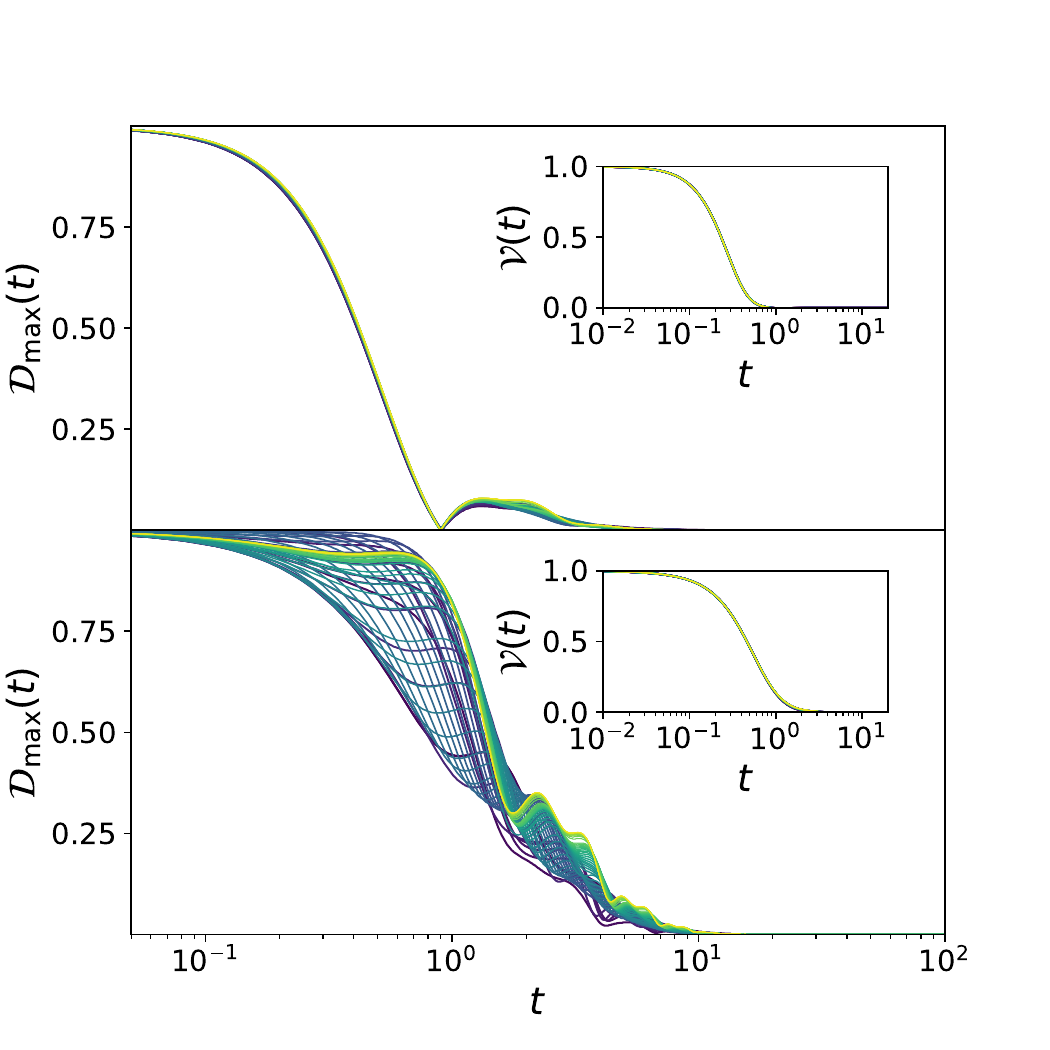}\caption{
Similar to \cref{fig:Dmax0.1} but with larger system--bath coupling constant $\eta=1$. }\label{fig:Dmax1} 
   \end{minipage}
\end{figure*}

\begin{figure*}[!htb]
   \begin{minipage}[t]{0.49\textwidth}
     \centering
     \includegraphics[width=1\linewidth]{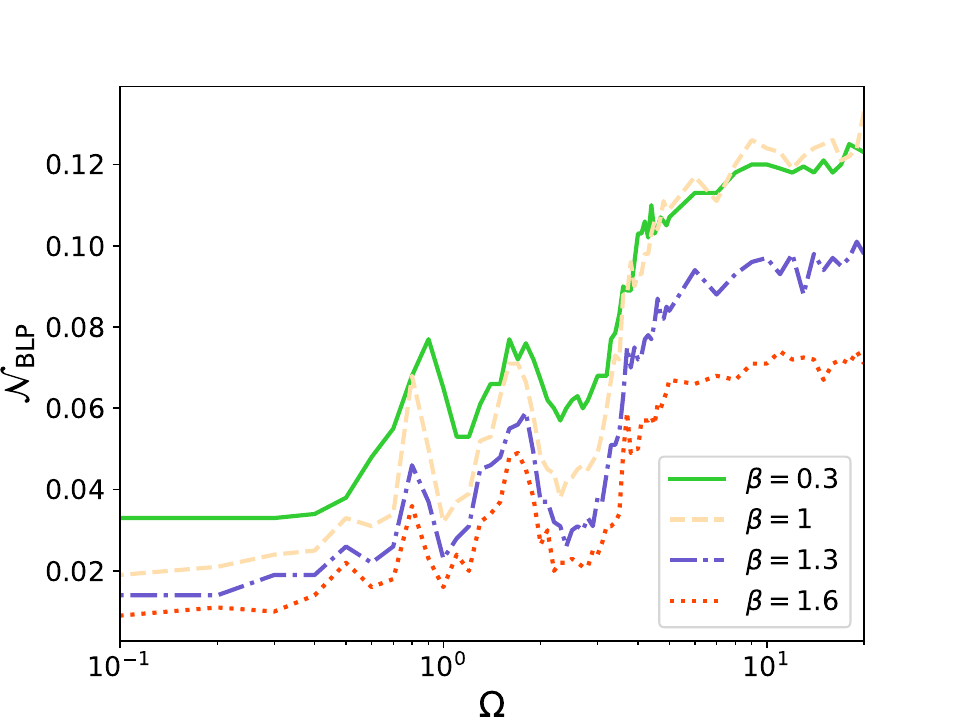}
     \caption{Integral measure of non-Makrovianity $\mc N_{\mr{BLP}}$ [\cref{eq:NBLP}], shown for four values of the inverse temperature $\beta$ and for increasing values of the frequency of the drive $\Omega$. 
The other parameters are set to  $\varepsilon_0=0$, $\varepsilon_{\mr d}=1$, $\eta=0.1$ and $\omega_{\mr{c}}=1$. 
The maximization in \cref{eq:NBLPv} has been performed with $N_{\mr{MC}}=10^6$ Monte Carlo samples.
}\label{fig:BLPeta0.1}
   \end{minipage}\hfill
   \begin{minipage}[t]{0.49\textwidth}
     \centering
     \includegraphics[width=1\linewidth]{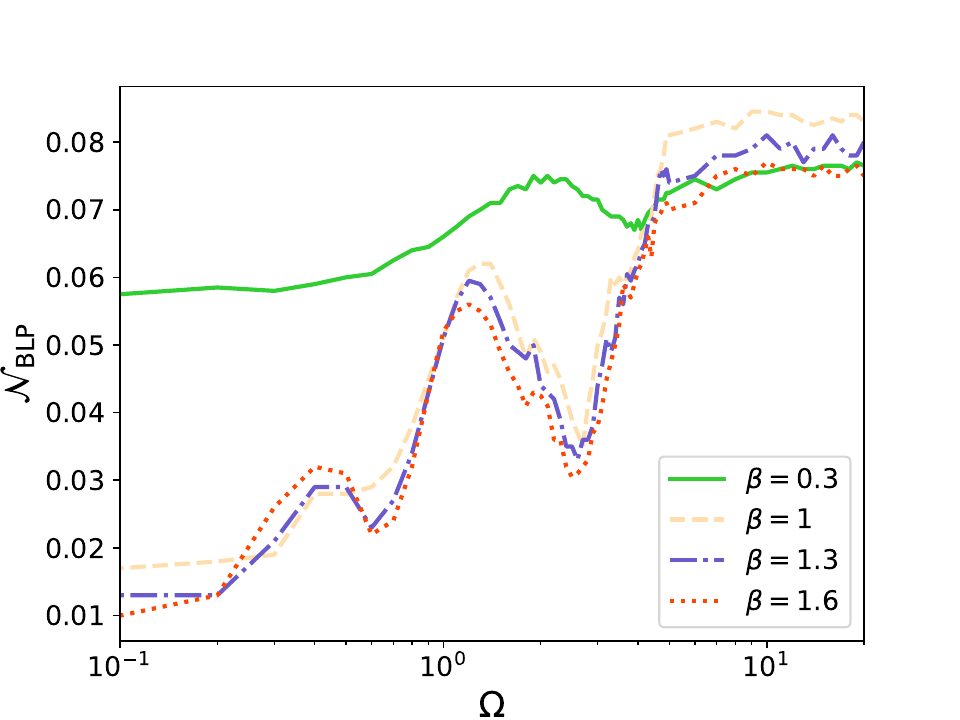}
     \caption{Similar to \cref{fig:BLPeta0.1} but with $\eta=1$ and $N_{\mr{MC}}=10^5$. }\label{fig:BLPeta1}
   \end{minipage}
\end{figure*}

Divisible time-dependent dynamical processes act as contractions of the trace distance, that is $\dot{\mc D}^{(1,2)}(t)\le 0$ for all $\bm v^{(1)}$ and $\bm v^{(2)}$. \cite{ruskai1994,breuer2016}
Non-divisibility can instead be interpreted as memory effects, given that time propagation on a finite time interval cannot be arbitrarily decomposed at intermediate times. 
The present observations led Breuer, Laine and Piilo (BLP) to introduce as a witness of non-Markovianity the integral measure
\begin{align}\label{eq:NBLP}
\mc N_{\mr{BLP}}  &= \max_{\hat\rho_{\mr s}^{(1)},\hat\rho_{\mr s}^{(2)} } \int_{\dot{\mc D}^{(1,2)}(t)>0} \mr d t\; \dot{\mc D}^{(1,2)}(t).
\end{align}
The maximization in \cref{eq:NBLP} guarantees that $\mc N_{\mr{BLP}}$ is exclusively a function of the dynamical map $\Phi_t$ describing the open quantum dynamics, rather than of any specific couple of initial densities.
It has been proven in Ref.~\onlinecite{wissmann2012} that the densities maximizing \cref{eq:NBLP} are mutually orthogonal and lie on the boundary of the  Hilbert space of the subsystem $\mc H_{\mr s}$.
For the specific case of two-level systems, those densities are represented by antipodal vectors of unit norm on the surface of the Bloch sphere.
We can hence rewrite \cref{eq:NBLP} as
\begin{equation}
\mc N_{\mr{BLP}}  = \max_{|\bm v|=1} \int_{\dot{\mc D}^{\bm v}(t)>0} \mr d t\; \dot{\mc D}^{\bm v}(t),\label{eq:NBLPv}
\end{equation}
where 
\begin{equation}
{\mc D}^{\bm v}(t)= \left\{\sum_{j=x,y,z} \left[\sum_{i=x,y,z}v_i\mc C_{ij}(t)\right]^2\right\}^{1/2}. \label{eq:Dv}
\end{equation}
The time evolution of the trace distance $\mc D_{\mr{max}}(t)$ maximizing \cref{eq:NBLPv}
is shown in \cref{fig:Dmax0.1,fig:Dmax1}.
Each figure corresponds to a fixed value of the system--bath coupling ($\eta=0.1$ and $\eta=1$, respectively).
Results are given for two values of the inverse temperature, $\beta=0.3$ and $\beta=1.6$ in the upper and lower panels, respectively. 
As in \cref{fig:Cmn}, the color scheme denotes increasing values of the frequency of the external drive $\Omega$ from blue/dark for $\Omega=0$ up to yellow/light for $\Omega=20$, with increasing warmth. 
The other system parameters are set to $\varepsilon_0=0$,  $\varepsilon_{\mr d}=1$ and $\omega_{\mr c}=1$.
By increasing the inverse temperature we observe an overall slower decay of the trace distance over time, as expected due to the decrease of thermal fluctuations of the environment leading to faster loss of coherence in the qubit.
Moreover, the time evolution of $\mc D_{\mr{max}}(t)$ appears to be highly sensitive to modulations of the driving frequency $\Omega$. 
To quantify the dependence of the information flow on the driving frequency, we show in \cref{fig:BLPeta0.1,fig:BLPeta1} phase diagrams of $\mc N_{\mr{BLP}}$ as a function of $\Omega$ for the same parameters of \cref{fig:Dmax0.1,fig:Dmax1} and two additional values of $\beta$.
The maximization in \cref{eq:NBLPv} is achieved by uniform Monte Carlo (MC) sampling of antipodal Bloch vectors on the surface of the Bloch sphere. \cite{clos2012} 
Satisfactory convergence is achieved with number of $N_{\mr {MC}}=10^6$ samples for \cref{fig:BLPeta0.1} and $N_{\mr {MC}}=10^5$ samples for \cref{fig:BLPeta1}.
Interestingly, revivals of information are observed for all considered systems.
This holds true also in the high-temperature regime at $\beta=0.3$, where the quantum environmental bath can be approximated by its classical counterpart (see \cref{subsec:dyn_ergo}).
Our analysis aligns with findings from other studies, witnessing quantum revivals even in presence of a classical environment. \cite{bordone2012,xu2013} 

In our results we detect peaks of $\mc N_{\mr{BLP}}$ for particular values of the drive linked to resonant revivals of information.  \cite{shuang2000}
This analysis indicates that, compared to the non-driven dynamics at $\Omega=0$, the inclusion of a periodic external field can effectively mitigate dissipation of information to the environment.
Moreover, this study reinforces the well-established significance of trace-distance approaches as valuable tools within the framework of quantum control. \cite{liu2011,chen2021}

In the above discussion on $\mc N_{\mr{BLP}}$ we pointed out that maximally distinguishable states of a qubit lay on the surface on the Bloch sphere.
This class of states maximizes also another scalar quantity, the volume underneath the surface of the sphere.
In the next section we discuss another approach to non-Markovianity based on this concept.


\subsubsection{Volume of accessible states}\label{subsubsec:volume}

From the propagation of the Bloch vector in \cref{eq:bloch_t} we  identify the volume on the Bloch sphere
\begin{equation}\label{eq:defV}
\mc V(t) = \lvert\det \mc C^T(t)\rvert
\end{equation}
accessible to the qubit at a given time $t$. \cite{lorenzo2013volume,chruscinski2017,puthumpally2018,mangaud2018}
Another insightful expression for $\mc V(t)$ follows from the first-order expansion 
\begin{align}
\mc V (t+\delta t) 
& =\llv \det\left[ \mc C^T(t) +\delta t \dot{\mc C}^T(t)\right] \rrv+ \mc O(\delta t^2) \nn \\
&=\llv \left\{\mc I_4  +  \delta t \tr_{\mr s} [\Xi(t)]\right\} \mc V(t) \rrv + \mc O(\delta t^2),\label{eq:incr_detM}
\end{align}
where we introduced the \emph{damping matrix} \cite{hall2014}
\begin{equation}\label{eq:XiC}
\Xi(t) = \dot {\mc C}^T(t) [\mc C^T(t)]^{-1} = \frac {\mr d}{\mr d t}\ln [\mc C^T(t)].
\end{equation}
\Cref{eq:incr_detM} leads to the
differential equation  
\begin{equation}
\frac{\mr d}{\mr d t}\mc V(t) = \tr_{\mr s}[ \Xi(t)]\mc V(t),
\end{equation}
solved by
\begin{equation}\label{eq:V_Xi}
\mc V(t) = \mc V (0)\exp\left\{\int_0^t\mr d t'\; \tr_{\mr s}[ \Xi(t')]\right\}.
\end{equation}
$\Xi(t)$ is also the generator of the dynamics in the Bloch representation. 
In fact, from \cref{eq:bloch_t}, 
\begin{align}
\dot v_\mu(t) 
&= \Xi_{\mu\nu}(t) v_\nu(t).
\end{align}
Equivalently, given that 
\begin{align}
\dot v_\mu(t) 
&=\tr_{\mr s}\left[\Lambda_t[\hat\rho_{\mr s}(t)]\hat\sigma_\mu\right]
\nn\\
&=\frac 12 \tr_{\mr s}\left[\Lambda_t[\hat\sigma_\nu]\hat\sigma_\mu \right]v_\nu(t),\label{eq:v_Lambda}
\end{align}
we identify 
\begin{equation}\label{eq:Xi_Lambda}
\Xi_{\mu\nu}(t) = \frac 12 \tr_{\mr s}\left[\Lambda_t[\hat\sigma_\nu]\hat\sigma_\mu\right].
\end{equation}
It can be proven that for completely positive open quantum dynamics the volume $\mc V(t)$ decreases monotonically in time. \cite{wolf2008,raja2018} 
This applies in particular to the BM limit discussed in \cref{sec:diss_drive}. 
Therefore, similarly to \cref{eq:NBLP}, it has been proposed in Ref.~\onlinecite{lorenzo2013} as a witness of non-Markovianity the integral measure
\begin{equation}\label{eq:NV}
\mc N_\mc V =\frac 1 {\mc V(0)} \int_{\dot {\mc V}(t)>0}\mr d t\;\dot{\mc V}(t).
\end{equation}
The time evolution of $\mc V(t)$ is shown in the  insets of \cref{fig:Dmax0.1,fig:Dmax1}.
We note that the timescale to relaxation of $\mc V(t)$ increases for larger inverse temperature $\beta$ and for smaller system--bath coupling $\eta$.
This suggests that the relaxation of the volume provides a simple and effective measure of decoherence, sensitive to the strength of dissipation induced by the bath.
However, for all considered system parameters $\mc V(t)$ decreases monotonically in time, hence $\mc N_{\mc V}=0$.
This result appears in contrast with the analysis of the trace distance, which, as discussed in \cref{subsubsec:TD}, witnesses non-Markovian effects in all the considered systems. 
A well-known issue of the volume approach is that $\mc V(t)$ accounts exclusively for the average dissipation over all quantum channels, potentially leading to an oversight of weak non-Markovian effects localized on a subset of channels on the Hilbert space. \cite{hall2014,chruscinski2017,mangaud2017}
We will extensively discuss this issue in the next \cref{subsubsec:canonical},
within the framework of the time-local master equation \cref{eq:canonical}.
Despite the aforementioned limitations, the analysis $\mc V(t)$ can still offer valuable insight into the invertibility of open quantum dynamics, a necessary condition to construct the time-local canonical master equation \cref{eq:canonical}. 
In particular, from the insets of \cref{fig:Dmax0.1}, we observe that, in the small coupling limit, the monochromatic drive can extend invertibility to longer times compared to the non-driven system with $\Omega=0$.

\subsubsection{Canonical rates}\label{subsubsec:canonical}

Studying the dynamics of the canonical rates within the time-local master equation \cref{eq:canonical} provides a rigorous method for quantifying non-Markovianity in open quantum systems.
In particular, it can be shown that, in the BM limit of \cref{eq:GQME_RDM}, the operators $\hat L_i$'s become time-independent, and the rates $\gamma_i$'s converge to non-negative constants. \cite{lindblad1976,devega2017} 
Beyond the BM approximation, the rates remain non-negative functions of time in case of completely positive divisible dynamics, as proven by the \emph{Gorini--Kossakowski--Sudarshan theorem}. \cite{gorini1976} 
In light of these observations, it has been proposed to measure non-Markovian effects on each channel by studying the negativity of the respective rate. \cite{hall2014}
This analysis provides in general deeper insight compared to the study of the volume of accessible states [\cref{eq:defV}]. \cite{hall2014,chrushinki2014}
In fact, the exponent of \cref{eq:V_Xi} can be rewritten as \cite{hall2014}
\begin{equation}\label{eq:Xi_gamma}
\tr_{\mr s}[\Xi(t)] = -2\sum_{i=1,2,3}\gamma_i(t).
\end{equation}
In conjunction with \cref{eq:V_Xi}, \cref{eq:Xi_gamma} tells us that $\mc V(t)$ provides information only on the average dissipation occurring on all quantum channels.
The analysis of all rates avoids instead possible compensation effects, which can in particular impair the detection of weak non-Markovian effects.

The knowledge of the full set of STCF for a given system provides complete information required to calculate the canonical dissipation rates.
In fact, the $\gamma_i(t)$'s are the eigenvalues of the \emph{decoherence matrix} $\xi(t)$, with components
\begin{align}
\xi_{ij}(t) &= \frac 14\tr_{\mr s}\left[\hat \sigma_\lambda\hat\sigma_i \Lambda_t[\hat\sigma_\lambda]\hat \sigma_j\right]\nn\\
&= \frac 14\tr_{\mr s}\left[\hat \sigma_\lambda\hat\sigma_i \hat\sigma_\rho\hat \sigma_j\right] \Xi_{\rho\lambda}(t),\label{eq:xi}
\end{align}
related to $\mc C(t)$ via \cref{eq:XiC}.
A proof of \cref{eq:xi} is shown in \cref{app:COS_superop}.
\begin{figure*}
\centering\includegraphics[width=0.73\linewidth]{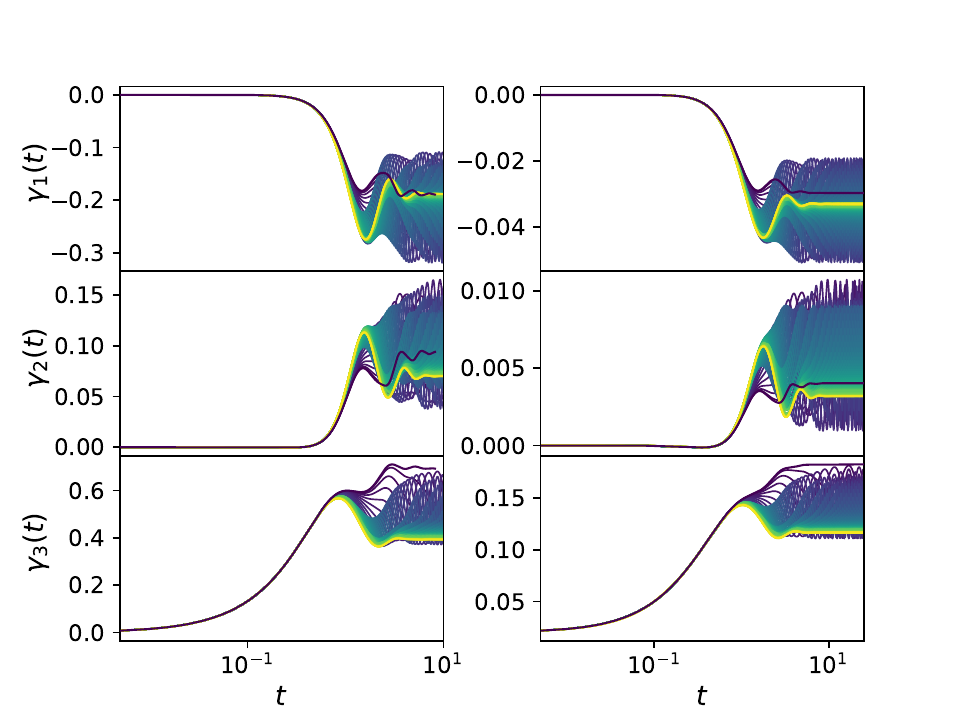}\caption{ Time evolution of the canonical rates. 
Each column corresponds to a given value of the  inverse temperature, $\beta=0.3$ and $\beta=1.6$ on the left and right columns, respectively.
Results are shown for increasing values of the driving frequency $\Omega$, with the same color scheme utilized in the previous figures. 
The other parameters are set to $\varepsilon_{0}=1$, $\varepsilon_{\mr d}=1$, $\eta=0.1$ and $\omega_{\mr c}=1$.}\label{fig:gamma}
\end{figure*}

An important caveat of this approach is that a solution of the rates at a given time $t$ is possible only provided the matrix $\mc C(t)$ is invertible.
However, for the systems considered in this work, the determinant \cref{eq:defV} vanishes at long times.
This can be inferred from the fact $\mc V(t)$ depends only on terms $\{\mc C_{ij}(t)\}_{ij}$, for $i,j\ge 1$ which vanish at long times (see the discussion in \cref{subsec:dyn_ergo}).
To account for this divergence, we calculated the solution of the rates up to the lowest threshold time $\tau_{\mr{th}}$ defined such that $\mc V( \tau_{\mr{th}})=10^{-3}$. 

The time evolution of the three canonical rates for the dissipative--driven qubit is given in \cref{fig:gamma}, for increasing values of the driving frequency from $\Omega=0$ (blue/dark) to $\Omega=20$ (yellow/light).
Each column corresponds to a fixed value of the inverse temperature ($\beta=0.3$ and $\beta=1.6$, at left and right, respectively).
The other system parameters are fixed to $\varepsilon_{0}=1$, $\varepsilon_{\mr d}=1$, $\eta=0.1$ and $\omega_{\mr c}=1$.
We observe that at both temperatures the lowest rate is negative at all times, and that it relaxes to a nonzero long-time plateau. 
The negativity of at least one quantum rate for all times is known in the literature as \emph{eternal non-Markovianity}. \cite{hall2014,megier2017,vaishy2022}
Remarkably, we observe for $\gamma_1(t)$ at $\beta=1.6$ (upper-right panel) that driving can increase the negativity of the rates at long times compared to the non-driven case (shown as the dark blue curve for $\Omega=0$).

Similarly to the study of the trace distance, the present analysis provides a well-defined criterion to measure the strength of non-Markovian effects.
Those serve as a useful proxy for  dissipation induced by the environment.
By increasing the system--bath coupling (to $\eta\gtrsim 1$) the rates would diverge, without displaying a stable non-Markovian negative plateau as observed in \cref{fig:gamma}.
Let us highlight that the positivity of ${\sum_{i=1,2,3}\gamma_i(t)}$ is the reason why non-Markovianity cannot be detected from the analysis of the accessible volume in \cref{subsubsec:volume}.  
 
As pointed out above, a drawback of the canonical-rates approach is that its applicability is restricted to invertible dynamical maps.
However, this limitation does not hinder the study of the time-nonlocal GQME, a general first-principled approach to quantum dissipation discussed in the next section.

\subsubsection{Generalized quantum master equation}\label{subsubsec:GQME}

In the following we introduce a dynamical approach to measure non-Markovian effects from the analysis of the GQME.

\begin{figure}
\centering\includegraphics[width=3.5in]{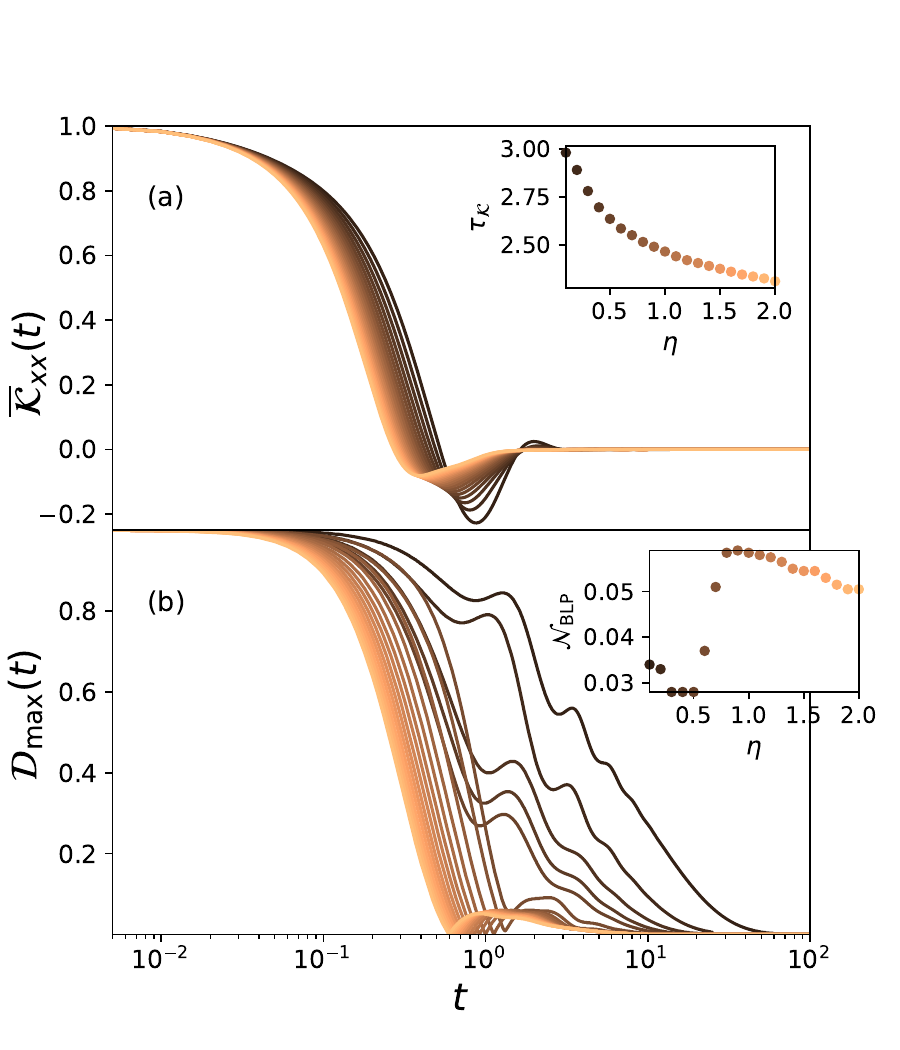}\caption{Panel (a): Diagonal component of the memory kernel \cref{eq:GQME_t_indep} and corresponding timescale to relaxation $\tau_{\mc K}$ [\cref{eq:tauK}].
Panel (b): time evolution of the maximal trace distance $\mc D_{\mr{max}}(t)$ and related integral measure of non-Markovianity $\mc N_{\mr{BLP}}$ [\cref{eq:NBLP}].
Results are shown for increasing values of the system--bath coupling from black/dark ($\eta=0.1$) to orange/light ($\eta=2)$. The other parameters of the system are fixed to $\beta=0.3$,  $\varepsilon_{0}=1$, $\varepsilon_{\mr d}=0$ and $\omega_{\mr c}=1$. }\label{fig:nonmarkovian}
\end{figure}

\Cref{eq:GQME_RDM1} can be rewritten in terms of an integro--differential equation of motion for the STCF's \cref{eq:Cmn}, given by
\begin{align}\label{eq:GQME}
\frac{\mr d}{\mr dt} \mc C(t,\tau) = \mc C(t,\tau)\mc X(t) - \int_{\tau}^t\mr d \tau' \; \mc C(\tau',\tau) \mc K(t,\tau').
\end{align}
The time-dependent drift $\mc X(t)$ and the memory-kernel matrix $\mc K(t,\tau)$ in \cref{eq:GQME} are defined in \cref{eq:X,eq:K} of \cref{app:GQME}, respectively.
In the same appendix we provide a derivation of \cref{eq:GQME}, together with details on how to calculate numerically the memory kernel from projection-free input time-correlation functions.
We consider here for simplicity the case of nondriven dynamics by fixing $\varepsilon_{\mr d}=0$ in \cref{eq:Hs}. 
In this case, \cref{eq:GQME} simplifies to \cref{eq:GQME_t_indep}.
Panel (a) of \cref{fig:nonmarkovian} shows the time evolution of the $\mu=\nu=x$ component of the memory kernel, normalized by its initial value, $\overline{\mc K}_{xx}(t)= {\mc K}_{xx}(t)/{\mc K}_{xx}(0)$.
Results are given for increasing values of the system--bath coupling from black/dark ($\eta=0.1$) to orange/light ($\eta=2)$. 
The other parameters are fixed to $\beta=0.3$,  $\varepsilon_{0}=1$ and  $\omega_{\mr c}=1$.
We observe that the kernel relaxes to zero faster for increasing values of $\eta$, converging towards the BM limit.
To quantify the deviation between the exact dynamics and the Markovian approximation of the GQME (see \cref{subsec:BM}), we measure a timescale to relaxation of the kernel defined by
\begin{equation}\label{eq:tauK}
\tau_{\mc K}(\delta) = \max_{\mu\nu} \left\{ \tau:\;\;\tfrac{\int_0^{\tau} \mr d t'\; |\mc K_{\mu\nu}(0)|}{\int_0^{+\infty} \mr d t'\; |\mc K_{\mu\nu}(0)|} = \delta \right\}, \hspace{3mm} \delta \in (0,1).
\end{equation}
In our numerical results, we fix $\delta=0.9$, and redefine $\tau_{\mc K} = \tau_{\mc K}(0.9)$.
As shown in the inset of panel (a) of \cref{fig:nonmarkovian}, $\tau_{\mc K}$ decreases monotonically  for increasing values of $\eta$.
By interpolating the decay of this curve for larger values of the coupling, it is possible to assess a threshold value for the system--bath coupling above which the BM equation provides a satisfactory description of the dynamics (see \cref{subsec:BM}).

In panel (b) of \cref{fig:nonmarkovian} we show the time evolution of the maximal trace distance, $\mc D_{\mr{max}}(t)$, while the integral measure of non-Markovianity, $\mc N_{\mr{BLP}}$, [\cref{eq:NBLP}] is displayed as a function of $\eta$ in the respective inset.
A steady increase in information revival occurs for intermediate system--bath coupling $0.5\lesssim \eta\lesssim 1$, at values corresponding to the transition from the coherent to the incoherent regime for the spin--boson model.
By further increasing the coupling constant ($\eta \gtrsim 1$) we observe a monotonic decrease of 
 information due to the strong dissipative effects exerted by the environment.
Both approaches discussed here tend converge towards a Markovian behavior for strong coupling with the environment.
Similarly to the study carried out in \cref{subsubsec:TD}, examining the trace distance offers a well-defined metric to quantify dissipation in the system.   
Furthermore, the complementary GQME approach can be effectively utilized to construct a Markovian approximation of the open dynamics in the strong-coupling regime, where revivals of information become negligible or vanish altogether.

\section{Conclusions}

In this paper we studied the open quantum dynamics of a dissipative--driven qubit, from multiple perspectives and with several approaches.
We integrated the HEOM of the model to propagate the RDM of the qubit over time across a wide range of parameters.
By decomposing the solution of the RDM  onto a complete set of correlation functions, we observed that the non-unital part of the dynamics leads to the emergence of non-equilibrium stationary states, involving the same periodicity of the total Hamiltonian. 

We calculated a witness of non-Markovianity by measuring the time evolution of the trace distance between two maximally distinguishable RDM's.
This approach provides a dynamical measure of information flow between the qubit and the external environment.
By tuning the frequency of an periodic monochromatic field, we witnessed interplay between dissipation and drive, leading to pronounced resonant peaks in information backflow.
This study highlights a valuable approach for devising control approaches to alleviate loss of coherence in dissipative quantum dynamics. \cite{morazotti2023,codyjones2012}

We noticed that for the model systems studied in this work the analysis of the volume of accessible states fails to witness weak signatures of non-Markovianity occurring on a finite subset of quantum channels.
This issue, as consistently observed in other studies, \cite{hall2014,chruscinski2017,mangaud2017} is due to the fact that the accessible volume only accounts for the average dissipation occurring over all quantum channels, and can overlook weak non-Markovian signatures.
The contradiction is resolved by decomposing the non-Hermitian part of the open quantum dynamics onto the canonical channels, and by measuring non-Markovian effects from the dissipation rates on each channel.
This numerical analysis allowed us to identify, for sufficiently small values of the system--bath coupling, stable signatures of \emph{eternal non-Markovianity}.\cite{gulacsi2023,hall2014,megier2017}
We observed that sufficiently strong drive can enhance non-Markovian effects compared to the non-driven dynamics.
This relevant observation could be effectively leveraged to advance quantum control strategies.

The time-nonlocal GQME is an exact equation of motion for the RDM of an open quantum system.
It provides an interpretation of dissipation in terms of dynamical memory effects, captured by fast-decaying memory kernels.
In this paper we introduced and studied a timescale to relaxation of the kernel as a measure of non-Markovian effects.
We showed that, for our model system and parameter domain, dynamical memory effects in the GQME decrease monotonically for increasing values of the system--bath coupling. 
Hence, the GQME formalism can be proficiently utilized as a numerical tool to construct an effective Markovian  approximation of open quantum dynamics for a strongly-dissipative bath. 

Several questions tackled in the present work are worth further investigation in future research, from both theoretical and technological perspectives.

It has been recently shown how features of eternal non-Markovianity emerge in the dynamics of superconducting qubits, a leading technology in the framework of quantum computing. \cite{gulacsi2023} 
It could be highly insightful to apply and extend the analysis of the present work to those systems, while tackling the challenging problem of improving coherence lifetime in multi-qubit gates.\cite{nakamura2024}

In this work we focused on the dissipative effects induced by the environment on a small quantum subsystem.
A complementary theoretical approach worth investigating involves a somehow reversed perspective, i.e. the study of how a quantum subsystem affects the dynamics of the environment in a thermal state.
This strategy could lead to meaningful insight, in particular in the development of QEC protocols relying on the design of engineered noise as a control resource.\cite{kapit2017}

The formalism of the time-local master equation is based on the decomposition of the open quantum dynamics onto a set of time-independent operators, associated to the canonical dissipation channels in the Hilbert space. 
It could be insightful to implement a dynamical and topological analysis of these channels to identify dissipation-free subspaces in the spirit of dynamical-decoupling techniques. \cite{fortunato2002} 

Among new perspectives for future research we mention the utilization of variational tools from control theory to design quantum systems subjected to minimal dissipation and stable coherence over long time scales. \cite{jirari2006,jirari2020,palao2008}
Finally, it can be highly insightful to connect the present analysis to  the process-tensor approach, as proposed in Ref.~\onlinecite{pollock2018,pollock2018nm}.
The method provides a robust operational criterion for measuring non-Markovian effects by examining the divisibility of quantum processes.
This strategy has proven effective in the investigation of non-Markovian noise in quantum-information devices, and in the development of successful error correction techniques based on dynamical decoupling. \cite{white2020,berk2023}

\section*{Data availability}

The data that support the findings of this study are available from the author upon reasonable request.


\section*{Acknowledgments}
The author would like to thank Tanja Schilling, Anja Seegebrecht, Fabian Koch, Heinz-Peter Breuer, Michael Thoss and Andreas Buchleitner for fruitful discussions and helpful comments. 
This project has been supported by the \emph{Deutsche Forschungsgemeinschaft} (DFG) via the Research Unit FOR5099 ``Reducing complexity of nonequilibrium'' (project no. 431945604).

\begin{appendix}

\section{High-frequency limit}\label{app:rotframe} 
In this appendix we determine a factorization of the propagator in \cref{eq:U} which allows us to straightforwardly calculate the $\Omega\to +\infty$ limit of the dynamics. 

Firstly let us remark that, due to the non-diagonal term $\Delta\hat\sigma_x$ in the total Hamiltonian $\hat H(t)$,
it is not possible to remove the time ordering in the integration in \cref{eq:U}. 
We can, however, factorize $\hat U(t)$ in two terms whose $\Omega\to\infty$ limit can be straightforwardly calculated. 
This is accomplished by introducing the rotating frame (R), defined by the unitary transformation
\begin{align}
\hat U_{\mr R}(t) &= \exp_+\left\{-i\hat\sigma_z\varepsilon_{\mr d} \int_0^t\mr dt'\; \cos(\Omega t')\right\}\nn\\
&= \hat {\mc I}_{\mr s} \cos \left[\frac {\varepsilon_{\mr d} \sin(\Omega t)}{\Omega} \right] - i \hat\sigma_z \sin  \left[\frac {\varepsilon_{\mr d} \sin(\Omega t)}{\Omega} \right], \label{eq:Ut}
\end{align}
rotating a quantum state $\ket {\psi(t)}$ in phase with an external drive.
The rotated state $ \ket {\psi_{\mr R}(t)}= \hat U_{\mr R}^\dagger(t) \ket {\psi(t)}$ obeys the Schr\"odinger equation \cite{shirai2018}
\begin{subequations}
\begin{align}\label{eq:schr_psiR}
i \frac{\partial}{\partial t}\ket{\psi_{\mr R}(t)} 
&= \left[\hat H_{\mr R}(t) - (\hat\sigma_z\otimes \hat{\mc I}_{\mr b})\varepsilon_{\mr d}\cos(\Omega t)\right]\ket{\psi_{\mr R}(t)},
\end{align}
\end{subequations}
where 
\begin{align}
\hat H_{\mr R}(t) 
&= \hat U_{\mr R}^\dagger(t)\hat H(t) \hat U_{\mr R}(t)\nn\\
&= [\varepsilon_0 +\varepsilon_{\mr d}\cos(\Omega t)]\hat\sigma_z \otimes \hat{\mc I}_{\mr b}+ \hat{\mc I}_{\mr s}\otimes\hat H_{\mr b} + \hat H_{\mr{sb}}  \nn\\
&\quad  + \Delta\Bigg[\cos \left(\frac {2\varepsilon_{\mr d} \sin(\Omega t)}{
	\Omega}\right)\hat\sigma_x\nn\\
&\quad - i\sin\left(\frac {2\varepsilon_{\mr d} \sin(\Omega t)}{\Omega}\right)\hat\sigma_y\Bigg]\otimes \hat{\mc I}_{\mr b}.
\end{align}
Let us introduce now the propagator
\begin{equation}
\hat V_{\mr R}(t) = \exp_+\left\{-i \int_0^t\mr d t'\; \left[\hat H_{\mr R}(t) - (\hat\sigma_z\otimes \hat{\mc I}_{\mr b})\varepsilon_{\mr d}\cos(\Omega t)\right]\right\},
\end{equation}
solving the Schr\"odinger equation \cref{eq:schr_psiR}.
Given that, for all $\ket{\psi(0)}$,
\begin{align}
\ket{\psi(t)} &= \hat U(t)\ket{\psi(0)}  =\hat U_{\mr R}(t)\ket{\psi_{\mr R}(t)}=\hat U_{\mr R}(t)\hat V_{\mr R}(t)\ket{\psi(0)},
\end{align}
it follows that 
\begin{equation}\label{eq:U_UR_VR}
\hat U(t)=\hat U_{\mr R}(t)\hat V_{\mr R}(t).
\end{equation}
\Cref{eq:U_UR_VR} allows us to calculate  the strong-frequency limit of the driven dynamics
\begin{align}\label{eq:Omega_to_inf}
\lim_{\Omega\to+\infty} \hat U(t)  &= \lim_{\Omega\to+\infty} \hat V_{\mr R}(t) = \lim_{\varepsilon_{\mr d}\to 0} \hat U(t).
\end{align}
\Cref{eq:Omega_to_inf} proves the correspondence between the limits $\Omega\to+\infty$ and $\varepsilon_{\mr d}\to 0$ discussed throughout the paper (see e.g. \cref{fig:Cmn}).

\section{GQME for dissipative--driven dynamics}\label{app:GQME}

\subsection{Derivation}\label{subsec:gen_GQME}

The Nakajima--Zwanzig GQME has affirmed itself as a powerful theoretical tool within the framework of open quantum systems and nonadiabatic dynamics. \cite{nakajima1958,zwanzig1960, brian2021, montoya2016,kelly2015}
Thus far, the approach has been extensively utilized to analyze and interpret non-Markovian effects in model systems described by time-independent Hamiltonians.
The formalism can be straightforwardly extended to the case of dissipative--driven dynamics, where the Hamiltonian involves an explicit  dependence on time.
In the present appendix we provide a derivation of the approach in this general case.
The following derivation applies specifically to a two-level subsystem, although it can be extended with no limitations to an arbitrary number $N$ of quantum levels (see the concluding remark of \cref{sec:diss_drive}).
We follow here the derivation originally presented in Ref.~\onlinecite{haake1973}, with the main difference that in our case we define projection operators in the spin basis.
The present approach is particularly well suited to tackle the analysis of the spin--spin correlation functions in a variety of models in quantum nonadiabatic dynamics.
\\

Let us consider a time-independent projection superoperator, $\mc P=\mc P^2$, acting on the total Hilbert space of system and environment.
Its complementary is defined by $\mc Q = \mathbb 1 - \mathcal P$, where $\mathbb 1$ is the identity superoperator. 

We aim to derive a non-Markovian equation of motion for the open quantum dynamics of correlation functions of the subsystem.
To do so, we start by decomposing the equation of motion of the two propagator onto the two directions of projection:
\begin{align}
&\frac{\mr d}{\mr d t}\exp_-\left\{i\int_{t_0}^t\mr d t'\; \mc L_{t'}\right\} 
=  i\exp_-\left\{ i\int_{t_0}^t\mr d t'\; \mc L_{t'}\right\}\mc P \mc L_t \nn\\
&\quad+i\exp_-\left\{i\int_{t_0}^t\mr d t'\; \mc L_{t'}\right\}\mc Q \mc L_t.\label{eq:P_Q}
\end{align}
We can now make use of the Dyson identity \cite{montoya2016}
\begin{align}
&\exp_-\left\{\int_{t_0}^t\mr d t'\; \mc B_{t'}\right\}
\nn\\
&=\exp_-\left\{ -\int_{t_0}^t\mr d t'\; \mc A_{t'}\right\}
 - \int_{t_0}^t\mr d \tau\; \exp_-\left\{ -\int_{t_ 0}^{\tau}\mr d t'\; \mc A_{t'}\right\} \nn\\
&\quad\times (\mc A_\tau+\mc B_\tau)\exp_-\left\{\int_{\tau}^t\mr d t'\; \mc B_{t'}\right\},\label{eq:dyson}
\end{align}
defined for two generic superoperators $\mc A_t$ and $\mc B_t$.
By replacing $\mc A_t =-i\mc L_t$ and $\mc B_t = i\mc Q \mc L_t$, we obtain 
\begin{align}
&\exp_-\left\{i\int_{t_0}^t\mr d t'\; \mc L_{t'}\right\} \nn\\
&=\exp_-\left\{ i\int_{t_0}^t\mr d t'\; \mc Q\mc L_{t'}\right\}+i \int_{t_0}^t\mr d \tau\; \exp_-\left\{ i\int_{t_0}^{\tau}\mr d t'\; \mc L_{t'}\right\} \nn\\
&\quad\times \mc P\mc L_{\tau}\exp_-\left\{i\int_{\tau}^t\mr d t'\; \mc Q\mc L_{t'}\right\},\label{eq:dyson_QL}
\end{align}
which can be replaced onto the right-hand side of \cref{eq:P_Q} to find \cite{kelly2016}
\begin{align}
&\frac{\mr d}{\mr d t}\exp_-\left\{i\int_{t_0}^t\mr d t'\; \mc L_{t'}\right\} = i\exp_-\left\{i\int_{t_0}^t\mr d t'\; \mc L_{t'}\right\}\mc P \mc L_t\nn\\
&\quad+i\exp_-\left\{i\int_{t_0}^t\mr d t'\; \mc Q\mc L_{t'}\right\}\mc Q \mc L_t\nn\\
&\quad-\int_{t_0}^t\mr d \tau\; \exp_-\left\{i\int_{t_0}^{\tau}\mr d t\; \mc L_{t'}\right\}\nn\\
&\quad\times \mc P\mc L_\tau\exp_-\left\{i\int_{\tau}^t\mr d t'\; \mc Q\mc L_{t'}\right\}\mc Q \mc L_t.\label{eq:eom_prop}
\end{align}
To construct an equation of motion for the STCF \cref{eq:Cmn}, we consider a Redfield-type projection superoperator \cite{kelly2016} spanning the full Hilbert space of the subsystem.
This is written in Liouville space \cite{mukamel1999} as
\begin{equation}\label{eq:P}
\mc P = | \hat \sigma_\lambda \rrangle \llangle \hat \rho_0 \hat \sigma_\lambda |.
\end{equation}
After multiplying \cref{eq:eom_prop} from left and right by $\llangle \hat\rho_0\hat\sigma_\mu|$ and $|\hat\sigma_\nu\rrangle$ respectively, we obtain the GQME \cref{eq:GQME}, involving the memory kernel
\begin{align}
&\mc K_{\mu\nu}(t,\tau) = \llangle \hat\rho_0\hat\sigma_\mu |\mc L_{\tau}\exp_-\left\{i \int_\tau^t\mr d t'\; \mc Q\mc L_{t'}\right\}\mc Q \mc L_t|\hat\sigma_\nu\rrangle \nn\\
& = \llangle \hat\rho_0\hat\sigma_\mu |\mc L_{\tau}\mc Q\exp_-\left\{i \int_\tau^t\mr d t'\;\mc Q\mc L_{t'}\right\}\mc Q \mc L_t|\hat\sigma_\nu\rrangle,\label{eq:K}
\end{align}
and drift matrix 
\begin{equation}\label{eq:X}
\mc X_{\mu\nu}(t)  = i \llangle \hat\rho_0\hat\sigma_\mu| \mc L_t|\hat\sigma_\nu\rrangle.
\end{equation}
The present choice of projection operators allows for a convenient simplification in the structure of the open quantum dynamics.
In fact, the matrix elements of the complicated projected dynamics in the second term of the right-hand side of \cref{eq:eom_prop} vanish identically:
\begin{align}
&i\llangle \hat\rho_0\hat\sigma_\mu |\exp_-\left\{i\int_{t_0}^t\mr d t'\; \mc Q\mc L_{t'}\right\}\mc Q \mc L_t| \hat\sigma_\nu\rrangle \nn\\
&=i\llangle \hat\rho_0\hat\sigma_\mu |\mc Q\exp_-\left\{i\int_{t_0}^t\mr d t'\; \mc L_{t'}\mc Q\right\}\mc Q \mc L_t| \hat\sigma_\nu\rrangle\nn\\
&=i\llangle \hat\rho_0\hat\sigma_\mu |\exp_-\left\{i\int_{t_0}^t\mr d t'\; \mc L_{t'}\mc Q\right\}\mc Q \mc L_t| \hat\sigma_\nu\rrangle  \nn\\
&\quad -i\llangle \hat\rho_0\hat\sigma_\mu |\hat\sigma_\lambda\rrangle \llangle\hat\rho_0\hat\sigma_\lambda|\exp_-\left\{i\int_{t_0}^t\mr d t'\; \mc L_{t'}\mc Q\right\}\mc Q \mc L_t| \hat\sigma_\nu\rrangle =0,
\end{align}
where we made use of the orthogonality of the scalar product $\llangle\hat\rho_0\hat\sigma_\mu |\hat\sigma_\nu \rrangle
=\tr[\hat\rho_0 \hat \sigma_\mu\hat\sigma_\nu]
=\delta_{\mu\nu}$.

The memory kernel in \cref{eq:K} depends on the
projected dynamics, which is hard to simulate numerically, both exactly and with approximated quasiclassical approaches. \cite{GQME} 
To circumvent this issue, we follow an established strategy originally introduced in Ref.~\onlinecite{shi2003}.
In particular, we construct a recursive relation for the memory kernel by inserting again the Dyson identity \cref{eq:dyson_QL} onto \cref{eq:K}: 
\begin{align}
&\mc K_{\mu\nu}(t,\tau) =  \llangle \hat\rho_0\hat\sigma_\mu |\mc L_{\tau}\mc Q\exp_-\left\{i \int_\tau^t\mr d t'\; \mc L_{t'}\right\} \mc Q\mc L_t|\hat\sigma_\nu\rrangle\nn\\
&\quad-i\int_{\tau}^t\mr d \tau'\; \llangle \hat\rho_0\hat\sigma_\mu |\mc L_{\tau}\mc Q\exp_-\left\{i\int_\tau^{\tau'}\mr d t'\; \mc L_{t'}\right\} |\hat\sigma_\lambda\rrangle\nn\\
&\quad\times \llangle \hat\rho_0\hat\sigma_\lambda|\mc L_{\tau'}\exp_-\left\{i \int_{\tau'}^t\mr d t'\; \mc Q\mc L_{t'}\right\}\mc Q\mc L_t|\hat\sigma_\nu\rrangle,\label{eq:aux_Kmu}
\end{align}
written in matrix form as
\begin{equation}\label{eq:volterra_K}
\mc K(t,\tau)= \mc K^{(1)}(t,\tau) +\int_\tau^t\mr d \tau'\; \mc K^{(3)}(\tau',\tau)\mc K(t,\tau'),
\end{equation}
where we introduced the two auxiliary kernels
\begin{subequations}\label{eq:auxK}
\begin{align}
\mc K_{\mu\nu}^{(1)}(t,\tau) &= \llangle \hat\rho_0\hat\sigma_\mu |\mc L_{\tau}\mc Q\exp_-\left\{i \int_\tau^t\mr d t'\; \mc L_{t'}\right\} \mc Q\mc L_t|\hat\sigma_\nu\rrangle,\label{eq:auxK1}\\
\mc K_{\mu\nu}^{(3)}(t,\tau) &= -i \llangle \hat\rho_0\hat\sigma_\mu |\mc L_{\tau}\mc Q\exp_-\left\{i \int_\tau^t\mr d t'\; \mc L_{t'}\right\}|\hat\sigma_\nu\rrangle.\label{eq:auxK3}
\end{align}
\Cref{eq:volterra_K,eq:auxK} allow us to conveniently calculate the solution of kernel \cref{eq:auxK} via projection-free input correlation functions.
\end{subequations}
In particular, by expanding the projectors in \cref{eq:auxK} we obtain the relations
\begin{subequations}
\begin{align}
\mc K^{(1)}(t,\tau) &= \frac{\mr d^2}{\mr d t\mr d \tau}\mc C(t,\tau) +\mc X(\tau)\frac{\mr d}{\mr d t}\mc C(t,\tau)\nn \\
&\quad-\frac{\mr d}{\mr d\tau}\mc C(t,\tau)\mc X(t) - \mc X(\tau) \mc C(t,\tau)\mc X(t),\label{eq:K1_der} \\
\mc K^{(3)}(t,\tau) &= \frac{\mr d }{\mr d \tau} \mc C(t,\tau)+\mc X(\tau)\mc C(t,\tau).\label{eq:K3_der}
\end{align}
\end{subequations}
By comparing \cref{eq:K1_der,eq:K3_der}, we find a relation between the two auxiliary kernels
\begin{equation}\label{eq:K1_K3}
\mc K^{(1)}(t,\tau) = \frac{\mr d}{\mr d t}\mc K^{(3)}(t,\tau)-\mc K^{(3)}(t,\tau)\mc X(t).
\end{equation}
Therefore, to obtain with projection-free input the full kernel \cref{eq:K} it suffices to calculate directly only $\mc K^{(3)}(t,\tau)$ via \cref{eq:Cmn}, and then to make use of the identities \cref{eq:K1_K3,eq:volterra_K}.

\subsection{Markovian approximation}\label{subsec:BM}

The memory kernel $\mc K(t)$ encapsulates complete information on the dynamical effects of the environment on the subsystem.
	Hence, one could ask at which extent the kernel relates to the other measures of non-Markovianity discussed in the present work.
	To tackle this question, we consider the Born--Markov (BM) approximation of \cref{eq:GQME}, \cite{OpenQuantum} corresponding to the assumption of memory-less dynamics $\mc K(t,\tau)\propto \delta(t-\tau)$.
	This leads to the time-local master equation
	\begin{subequations}\label{eq:GQME_BM}
		\begin{align}
			\frac{\mr d}{\mr d t }{\mc C}_{\mr{BM}}(t,\tau) &= \mc C_{\mr{BM}}(t,\tau)\mc M_{\mr{BM}}(t,\tau),\\
			\mc M_{\mr{BM}}(t,\tau) &= \mc X(t)-\int_\tau^t\mr d \tau'\; \mc K(t,\tau') ,
		\end{align}
	\end{subequations}
	which is solved by 
	\begin{subequations}
		\begin{align}
			\mc C_{\mr{BM}}(t,\tau) &= [\Phi_{\mr{BM}}(t,\tau)\mc I_4]^T ,\\
			\Phi_{\mr{BM}}(t,\tau) &= \exp_+ \left\{  \int_\tau^t\mr d t'\; \mc M_{\mr{BM}}^T(t',\tau) \right\},\label{eq:Phi_BM}
		\end{align}
	\end{subequations}
	where we noticed that the initial value $\mc C_{\mr{BM}}(0,0)$ is equal to the $4\times 4$ identity matrix $\mc I_4$.
	Note that the BM propagator $\Phi_{\mr{BM}}(t,\tau)$, being a time-ordered exponential, is divisible, \cite{evans2007} i.e., 
	\begin{equation}\label{eq:divisib}
		\Phi_{\mr{BM}}(t,0) = \Phi_{\mr{BM}}(t,\tau)\Phi_{\mr{BM}}(\tau,0),\hspace{3mm}
		0\le\tau\le t.  
	\end{equation}
	This indicates that the BM approximation of the GQME is consistent with the correspondence between Markovianity and divisibility of the dynamical map, laying at the foundation of all the approaches to quantum dissipation discussed in the present work. \cite{breuer2016,rivas2010,chruscinki2011,chrushinki2014} 
\\

For a non-driven time-independent Hamiltonian, \cref{eq:GQME} simplifies to 
	\begin{align}\label{eq:GQME_t_indep}
		\frac{\mr d}{\mr dt} \mc C(t) = \mc C(t)\mc X- \int_0^t\mr d \tau' \; \mc C(\tau') \mc K(t-\tau'),
	\end{align}
	with corresponding BM approximation
	\begin{subequations}\label{eq:GQME_t_indep_BM}
		\begin{align}
			\frac{\mr d}{\mr d t }{\mc C}_{\mr{BM}}(t) &= \mc C_{\mr{BM}}(t)\mc M_{\mr{BM}},\\
			\mc M_{\mr{BM}} &= \mc X-\int_0^{+\infty}\mr d \tau'\; \mc K(\tau'),
		\end{align}
	\end{subequations}
	generating divisible dynamics.
	One can prove that \cref{eq:GQME_t_indep_BM} preserves the same long-time limits of the exact dynamics \cref{eq:GQME_t_indep} (see Appendix B of Ref.~\onlinecite{GQME} for further details).
	This supports the relevance and physical significance of the present Markovian approximation of the GQME. 

\section{Canonical form of the time-local master equation}\label{app:canonical}
Studying the dynamics of open quantum systems via the time-local master equation \cref{eq:canonical} provides a rigorous and well defined strategy to witness the occurrence of non-Markovian effects.
In this appendix we present a detailed derivation of this master equation, following Refs.~\onlinecite{mangaud2017,hall2014,andersson2007}. 
\\

First of all, we observe that the linearity of the generator $\Lambda_t$ [\cref{eq:GQME_RDM}] implies that, if the propagator $\Phi_t$ is invertible, two sets of operators $\{\hat A_\mu(t)\}_\mu$ and $\{\hat B_\mu(t)\}_\mu$ exist such that
\begin{align}\label{eq:rhosAB}
\Lambda_t\hat\rho_{\mr s}(t) = \hat A_\mu(t)\hat\rho_{\mr s}(t) \hat B_\mu^\dagger(t).
\end{align}
By expanding 
\begin{equation}
\hat A_\mu(t) =  \hat\sigma_\nu a_{\nu\mu }(t) , \hspace{10mm}\hat B_\mu(t) =\hat\sigma_\nu b_{\nu\mu }(t)  , 
\end{equation}
we rewrite \cref{eq:rhosAB} as
\begin{align}
\Lambda_t\hat\rho_{\mr s}(t) &= a_{\nu\mu }(t) \hat\sigma_\nu\hat\rho_{\mr s}(t)  b^*_{\lambda\mu }(t) \hat\sigma_\lambda = \xi_{\nu \lambda}(t) \hat\sigma_\nu\hat\rho_{\mr s}(t) \hat\sigma_\lambda ,\label{eq:rhos_xi1}
\end{align}
where we introduced the \emph{decoherence matrix} $\xi(t)$, with components $\xi_{\nu \lambda}(t) =  a_{\nu\mu }(t)b^*_{\lambda\mu }(t)$ [see also \cref{eq:xi} for an alternative expression].
Given that $\hat\rho_{\mr s}(t)$ and $\Lambda_t\hat\rho_{\mr s}(t)$ are Hermitian operators, $\xi(t)$ is a Hermitian matrix at all times.
In fact,
\begin{align}
[\Lambda_t\hat\rho_{\mr s}(t)]^\dagger 
&=  \xi_{\nu\lambda}^*(t) \hat\sigma_\lambda\hat\rho_{\mr s}(t) \hat\sigma_\nu
= \xi_{\lambda\nu}^*(t) \hat\sigma_\nu\hat\rho_{\mr s}(t) \hat\sigma_\lambda,\label{eq:rhos_xi0.5}
\end{align}
and, by comparing \cref{eq:rhos_xi1,eq:rhos_xi0.5}, it follows that $\xi_{\nu \lambda}(t) =\xi_{\lambda \nu }^*(t)$.
To identify in \cref{eq:rhos_xi1} the Hermitian part of the open dynamics, we expand
\begin{align}
\Lambda_t\hat\rho_{\mr s}(t) &= \xi_{00}(t) \hat\rho_{\mr s}(t) + \xi_{i0}(t)\hat\sigma_i \hat\rho_{\mr s}(t) + \xi_{0j}(t)\hat\rho_{\mr s}(t) \hat\sigma_j \nn\\
&\quad + \xi_{ij}(t) \hat\sigma_i \hat\rho_{\mr s}(t)\hat\sigma_j,\label{eq:rhos_xi2}
\end{align}
and, by defining\begin{equation}
\hat h_{\mr c}(t) =\tfrac 12 \xi_{00}(t) \hat{\mc I}_{\mr s} + \xi_{i0}(t)\hat\sigma_i,
\end{equation}
we recast \cref{eq:rhos_xi2} as
\begin{align}
\Lambda_t \hat\rho_{\mr s}(t)  &= \hat h_{\mr c}(t)\hat\rho_{\mr s}(t) + \hat\rho_{\mr s}(t) \hat h_{\mr c}^\dagger(t) + \xi_{ij}(t) \hat\sigma_i \hat\rho_{\mr s}(t)\hat\sigma_j.\label{eq:rhos_xi3}
\end{align}
By taking the trace of \cref{eq:rhos_xi3} and making use of the trace preservation of the density matrix
 ($\tr_{\mr s}[\Lambda_t\hat\rho_{\mr s}(t)]=0$), we obtain
\begin{align}
\hat h_{\mr c}(t) + \hat h_{\mr c}^\dagger(t) = -\xi_{ij}(t)\hat\sigma_j \hat\sigma_i.
\end{align}
The Hermitian operator 
\begin{equation}\label{eq:Hc}
\hat H_{\mr c}(t) = \frac {i}2 \left[\hat h_{\mr c}(t)- \hat h_{\mr c}^\dagger(t)\right],
\end{equation}
generates the unitary part of the dynamics in \cref{eq:rhos_xi3}:
\begin{align}
\Lambda_t \hat\rho_{\mr s}(t)
&= -i[\hat H_{\mr c}(t),\hat\rho_{\mr s}(t)] \nn\\
&\quad +\xi_{ij}(t) \left[  \hat\sigma_i \hat\rho_{\mr s}(t)\hat\sigma_j -\tfrac 12  \left\{\hat\sigma_j\hat\sigma_i ,\hat\rho_{\mr s}(t)\right\}  \right].\label{eq:rhos_xi4}
\end{align}

The negativity of the eigenvalues of $\xi(t)$ provides a measure of the discrepancy from the Markovian limit of the dynamics.
In particular, we can decompose $\xi(t)$ in terms of its real  eigenvalues  $\gamma_i(t)$'s and corresponding orthonormal eigenvectors $\bm u^{(i)}(t)$'s according to
\begin{align}
\xi_{ij}(t) 
&= u_{i}^{(k)}(t) \gamma_{k}(t) \left[u^{(k)}_j(t)\right]^* = \mc U_{ik}(t)\gamma_k(t) \mc U_{ik}^\dagger(t),
\end{align}
where $\mc U_{ij}(t) = u_i^{(j)}(t)$.
The time-dependent operators
\begin{align}\label{eq:Lsigma}
\hat L_i(t) &= \mc U_{ij}(t)\hat\sigma_j,
\end{align}
allow us to rewrite \cref{eq:rhos_xi4} as
\begin{align}
\Lambda_t \hat\rho_{\mr s}(t) 
&=  - i [\hat H_{\mr c}(t),\hat\rho_{\mr s}(t)]\nn\\
&\quad + \left( \mc U^\dagger_{ki}(t)\xi_{ij}(t)\mc U_{jl}(t)\right)  \hat L_k(t) \hat\rho_{\mr s }(t)  \hat L^\dagger_l(t) \nn\\
&\quad -\frac 12 \left(\mc U^\dagger_{li}(t)\xi_{ij}(t)\mc U_{jk} (t)\right)  \{ \hat L^\dagger_k(t) \hat L_l(t) ,\hat\rho_{\mr s}(t)\}, 
\end{align}
which corresponds to \cref{eq:canonical}, given that $ \left[\mc U^\dagger(t)\xi(t) \mc U(t)\right]_{ij} = \delta_{ij}\gamma_j(t)$.

\section{Decomposition of superoperators of the subsystem}\label{app:COS_superop}

In the present appendix we include the derivation of \cref{eq:xi}, an identity relating the canonical rates to the damping matrix \cref{eq:XiC}.
The following proof corresponds to Lemma 2.2  Ref.~\onlinecite{gorini1976}. 
\\ 
\\
{\bf Lemma: }\textit{Let us denote by $\{\hat \pi_\mu\}_\mu$ a complete orthonormal set (COS) of operators on $\mc H_{\mr s}$.
For $\hat O:\mc H_{\mr s}\to \mc H_{\mr s}$, the family of superoperators $\{\mc G_{\mu\nu}\}_{\mu\nu}$ defined by 
\begin{equation}\label{eq:Gmn}
\hat O\mapsto \mc G_{\mu\nu}\hat O = \hat \pi_\mu\hat O\hat \pi_\nu^\dagger,
\end{equation}
is a COS in $L(\mc H_{\mr s})$, the linear space of superoperators acting on $\mc H_{\mr s}$.
}
\\
\\
{\bf Proof:} 
First of all, we notice that 
\begin{equation}\label{eq:piApi}
\hat \pi_\mu^\dagger \hat O \hat\pi_\mu = \hat{\mc I}_{\mr s} \tr_{\mr s}[\hat O].
\end{equation}
In fact, being the left-hand side of \cref{eq:piApi} invariant under a change of COS, 
we can replace $
\{\hat\pi_\mu\}_\mu$ with $\{\hat E_{ab} = \ket a \bra b\}_{a,b=0,1}$, and expand 
\begin{equation}
\hat E_{ab}^\dagger \hat O \hat E_{ab} = \ket b \braket{a|\hat O | a} \bra b = \hat {\mc I}_{\mr s}\tr[\hat O].
\end{equation}
$L(\mc H_{\mr s})$ is a unitary space 
with inner product
\begin{align}\label{eq:inner_superop}
\langle \mc A, \mc B\rangle = \tr_{\mr s}[(\mc A \hat\tau_\mu)^\dagger (\mc B \hat\tau_\mu)], \hspace{5mm}\mc  A, \mc B \in L(\mc H_{\mr s}),
\end{align}
also independent on any specific choice of COS $\{\hat \tau_\mu\}_\mu$. 
We can now prove that $\{\mc G_{\mu\nu}\}_{\mu\nu}$ is a COS in $L(\mc H_{\mr s})$ . 
In fact, via \cref{eq:piApi},
\begin{align*}
\langle{\mc G}_{\alpha\beta},{\mc G}_{\mu\nu} \rangle &= \tr_{\mr s}\left[({\mc G}_{\alpha\beta}\hat \tau_\lambda)^\dagger({\mc G}_{\mu\nu}\hat \tau_\lambda)\right] \nn\\
&= \tr_{\mr s}\left[\left(\hat\pi_\alpha\hat \tau_\lambda\hat \pi_\beta^\dagger\right)^\dagger\left(\hat\pi_\mu\hat\tau_\lambda\hat\pi_\nu^\dagger\right)\right]\nn\\
&= \tr_{\mr s}\left[\hat\pi_\beta (\hat \tau^\dagger_\lambda \hat \pi_\alpha^\dagger \hat\pi_\mu\hat\tau_\lambda)\hat\pi_\nu^\dagger\right]\nn\\
&= \tr_{\mr s}[\hat\pi_\alpha^\dagger \hat \pi_\mu ]\tr_{\mr s}[\hat\pi_\beta \hat\pi_\nu^\dagger]=\delta_{\alpha\mu}\delta_{\beta\nu}.\;\;\blacksquare
\end{align*}
\\

The above result implies that the generator $\Lambda_t$ can be decomposed on the basis $\{\mc G_{\mu\nu}\}_{\mu\nu}$ according to
\begin{equation}\label{eq:Lambdarhos_scal}
\Lambda_t\;\bullet  = \langle \Lambda_t, {\mc G}_{\nu\lambda}\rangle {\mc G}_{\nu\lambda}\;  
\bullet. 
\end{equation}
\Cref{eq:xi} follows by applying \cref{eq:Lambdarhos_scal} to $\hat\rho_{\mr s}(t)$, expressing both the scalar product and the superoperators $\mc G_{\mu\nu}$ in terms of the COS $\tfrac 1 {\sqrt 2}\{\hat\sigma_\mu\}_\mu$, and by comparing with \cref{eq:rhos_xi1}. 

\end{appendix}
\FloatBarrier
%

\end{document}